\documentclass[preprint]{aastex}
\pdfoutput=1

\usepackage{caption}
\usepackage{subcaption}
\usepackage{fixltx2e}
\usepackage[usenames,dvipsnames]{color}
\usepackage{amsmath}
\usepackage{multirow}
\usepackage{graphics}
\usepackage{txfonts}
\usepackage{enumitem}

\begin{document}


\title{High-energy particle transport\\in 3D hydrodynamic models\\of colliding-wind binaries}

\author{K.~Reitberger, R.~Kissmann, A.~Reimer, O.~Reimer}
\affil{Institut f\"ur Astro- und Teilchenphysik and Institut f\"ur Theoretische Physik, Leopold-Franzens-Universit\"at Innsbruck, A-6020 Innsbruck, Austria} 
\and 
\author{G.~Dubus}
\affil{UJF-Grenoble 1/CNRS-INSU, Institut de Plan\'{e}tologie et d'Astrophysique de Grenoble (IPAG), UMR 5274, Grenoble, France}
\email{klaus.reitberger@uibk.ac.at}


\begin{abstract}
Massive stars in binary systems (as WR140, WR147 or $\eta$ Carinae) have long been regarded as potential sources of high-energy $\gamma$-rays. The emission is thought to arise in the region where the stellar winds collide and produce relativistic particles which subsequently might be able to emit $\gamma$-rays.
Detailed numerical hydrodynamic simulations have already offered insight in the complex dynamics of the wind collision region (WCR), while independent analytical studies, albeit with simplified descriptions of the WCR, have shed light on the spectra of charged particles.
In this paper, we describe a combination of these two approaches. We present a 3D-hydrodynamical model for colliding stellar winds and compute spectral energy distributions of relativistic particles for the resulting structure of the WCR.
The hydrodynamic part of our model incorporates the line-driven acceleration of the winds, gravity, orbital motion and the radiative cooling of the shocked plasma. In our treatment of charged particles we consider diffusive shock acceleration in the WCR and the subsequent cooling via inverse Compton losses (including Klein-Nishina effects), bremsstrahlung, collisions and other energy loss mechanisms.
\end{abstract}



\section{INTRODUCTION} 
In recent years, several models of mostly analytical nature have addressed the question whether binary systems of massive stars without compact objects are liable to produce high-energy $\gamma$-ray emission \citep[e.g.,][]{Reimer2006,Pittard2006,Benaglia2003}.
These studies argue that such objects (i.e. systems containing a Wolf-Rayet (WR) star and an OB-type star) provide suitable environments for efficient particle acceleration and subsequent $\gamma$-ray emission. Electrons and protons are thought to be accelerated at the shock fronts tracing the edge of the regions where stellar winds collide with supersonic velocities. Several $\gamma$-ray emission mechanisms -- inverse Compton (IC) scattering, relativistic bremsstrahlung and $\pi^0$-decay -- can produce $\gamma$-rays at GeV and TeV energies with sufficient fluxes  to allow for detection with instruments as the \textit{Fermi}-Large Area Telescope (LAT) or perhaps even HESS, MAGIC and VERITAS.

In contrast to expectations, no such detection of $\gamma$-ray emission linked to colliding-wind binaries (CWBs) has been reported so far, with one notable exception: the highly unusual object $\eta$~Carinae is the only CWB system unambiguously linked to high-energy $\gamma$-ray emission \citep{Reitberger2012}. This system most likely
consists of a WR star and a high-mass Luminous Blue Variable (LBV). Both stars are enveloped in a huge dust and gas cloud, the Homunculus-nebula, which originated in a massive outburst of the LBV in the year of 1843. Although this source exhibits a number of unique characteristics, no explanation for its high $\gamma$-ray flux -- compared to the non-detection of other CWBs -- can be given so far.  
Dedicated observations of WR-OB systems as WR 147 or WR 140 (for which models predicted fluxes above LAT detection thresholds) have yielded upper limits \citep[][]{Werner2013}.

Detailed 3D hydrodynamical (HD) simulations \citep{Pittard2009} have recently explored the highly dynamical nature of the WCR in CWBs and its strong dependence on stellar and orbital parameters. The complex density, velocity and temperature structure of the colliding winds have further been used to model the thermal radio and X-ray emission in such systems \citep{Pittard2010,Pittard2010b}. 

This work aims at a numerical computation of the spectral energy distribution of charged particles within a numerical HD model of the WCR.  By solving a transport equation including spatial convection, diffusive shock acceleration (DSA) and various cooling processes for electrons and protons at every grid point of the HD simulation, we simulate the time-dependent 3D spatial distribution of particles at different energies. As cooling processes we take IC emission, bremsstrahlung, nucleon-nucleon interaction, adiabatic cooling and others into account. The particles are injected at the shock fronts of the wind collision region and subsequently gain energy by DSA.

We developed a code which -- by taking stellar, stellar wind and orbital parameters of a given binary system as input -- solves the 3D distribution of density, velocity and temperature fields of the wind plasma as well as the energy spectra for electrons and protons for every point on a numerical grid.
The resulting spatially varying particle spectra will serve as input for complementary studies, computing the components of ensuing $\gamma$-ray emission.

In Section 2 we introduce the numerical and HD setup that we use to simulate a 3D distribution of radiatively driven winds in a binary system. Our method of dealing with the spatial and energetic evolution of particle species via solving a transport equation is thoroughly discussed in Section 3. In Section 4 we present results obtained for a typical CWB system in terms of spatial and spectral distribution of high-energy electrons and protons. Section 5 provides a summary of our findings as well as an outlook on future developments.

\section{HYDRODYNAMICS} 
\subsection{Numerical Setup}
To simulate the hydrodynamics of the stellar winds we use the \textsc{Cronos} code \citep{Kissmann2008,Kleimann2009} which is a finite-volume magneto-hydrodynamic (MHD) code optimized for
the simulation of compressible astrophysical plasmas.  The code is
second-order accurate in space and allows for Cartesian, cylindrical,
and spherical grid layouts.  \textsc{Cronos} uses approximate Riemann
solvers for the time integration of the HD and MHD equations,
where the choice of Riemann solvers includes Harten-Lax-van Leer (hll, used in this study), hll-Contact (hllc) and hll-Discontinuities (hlld).
Time integration is done via a second- or third-order Runge-Kutta integrator, where the time discretisation utilises the semidiscrete approach.

In the present case, \textsc{Cronos} is used to solve the HD equations,
\begin{equation}
\label{euler1}
\frac{\partial\rho}{\partial t}+\vec{\nabla}\cdot(\rho\vec{v}) = 0
\end{equation}

\begin{equation}
\label{euler2}
 \frac{\partial\rho \vec{v}}{\partial t}+\vec{\nabla}\cdot(\rho \vec{v}\vec{v} + \hat{P}) = \rho \vec{f}
\end{equation}

\begin{equation}
\label{euler3}
\frac{\partial \epsilon}{\partial t}+\vec{\nabla}\cdot[(\epsilon + P)\vec{v}] = \left(\frac{\rho}{m_H}\right)^2\Lambda(T)+\rho \vec{f}\cdot\vec{v} 
\end{equation}
where the left-hand side of the equations is solved by \textsc{Cronos} intrinsically, whereas the right-hand side (i.e. the force term $\vec{f}$ and the radiative cooling term $\Lambda(T)$ are dealt with in dedicated additional modules as outlined below.

$\rho$ is the mass density, $\vec{v}$ is the velocity vector, $P$ the scalar pressure, $\vec{f}$ the sum of all external forces, $\epsilon = \frac{\rho}{2}v^2+e$ is the total energy per volume, $e$ the internal energy per volume and T the temperature.
We use the ideal gas equation of state $P=(\gamma-1)e$ with the adiabatic index for a monoatomic gas $\gamma = \frac{5}{3}$.

\subsection{The force term $\vec{f}$}

The force density term $\vec{f}$ consists of three components: gravity, radiative line acceleration due the ions in the wind and the radiative acceleration due to  photons scattering off electrons. It can be written as
\begin{equation}
\vec{f} = \sum_{i=1}^2\left(-GM_{\ast,i}\frac{\vec{r}_{i}}{r_i^3}+\vec{g}_{rad,i}^L+\vec{g}_{rad,i}^e\right)
\end{equation}
where the index $i$ indicates each star. The vector $\vec{r}_{i}$ is given relative to the star $i$. We assume radiative acceleration to be directed radially from the stars. Therefore, 
\begin{equation}
\vec{g}_{rad,i}^e=g_{rad,i}^e\frac{\vec{r}_{i}}{r_i} \qquad \mbox{with} \quad g_{rad,i}^e =\frac{\sigma_eL_{\ast,i}}{4\pi r_i^2c}
\end{equation}
where $\sigma_e$ is the specific electron opacity due to Thomson scattering and $L_{\ast,i}$ is the luminosity of star $i$. \\

Determining the line-acceleration requires an integration over the finite stellar disk (as a point source based approach would yield significantly erroneous results close to the star \citep[see][]{Cassinelli1999}. This integral can be simplified by assuming azimuthal symmetry of the stellar disk.
To approximate the contribution of the wide spectrum of optically thick and thin spectral lines of the ions in the wind, we rely on the standard Castor-Abbott-Klein (CAK) formalism first introduced by \citet{CAK} and later modified and improved by \cite{Pauldrach1986}. This approach replaces the required summation over all lines by a simple parametrization with two parameters $\alpha$ and $k$.
Taking all this into account, the line-acceleration term \citep[as given in][]{Gayley1997} can -- with some modification -- be expressed as follows:
\begin{equation}
\label{CAK}
\vec{g}_{rad,i}^L=g_{rad,i}^L\frac{\vec{r}_{i}}{r_i} \qquad \mbox{with} \quad g_{rad}^L =\frac{2 k\sigma_e^{1-\alpha}}{c}\frac{L_{\ast,i}}{4\pi R_{\ast,i}^2}  \int_0^{\theta_{\ast,i}} \left(\frac{\vec{n_i}\cdot\vec{\nabla}(\vec{n_i}\cdot\vec{v})}{\rho v_{th}}\right)^\alpha \sin\theta\cos\theta d\theta
\end{equation}
with $k$ and $\alpha$ being the CAK-parameters mentioned above.
$\rho$ is the mass density of the wind in the cell for which the acceleration is calculated. $v_\mathrm{th}$ is the thermal velocity of ionized hydrogen, the angle $\theta_{\ast,i}$ marks the edge of the stellar disk relative to the point for which we derive the acceleration. It is defined by $\sin\theta_{\ast,i}=\frac{R_{\ast,i}}{r_i}$.
The projected velocity gradient $\vec{n_i}\cdot\vec{\nabla}(\vec{n_i}\cdot\vec{v})$ depends on the unit vector towards a point on the stellar surface $\vec{n_i}$ depending again on $\theta$. In Euclidean coordinates it can be expressed as
\begin{equation}
\label{grad}
\begin{aligned}
\vec{n}\cdot\vec{\nabla}(\vec{n}\cdot\vec{v})=n_x^2\frac{\partial v_x}{\partial x}+n_y^2\frac{\partial v_y}{\partial y}+n_z^2\frac{\partial v_z}{\partial z}+n_xn_y\left(\frac{\partial v_y}{\partial x}+\frac{\partial v_x}{\partial y}\right)+\\+n_xn_z\left(\frac{\partial v_z}{\partial x}+\frac{\partial v_x}{\partial z}\right)+n_yn_z\left(\frac{\partial v_z}{\partial y}+\frac{\partial v_y}{\partial z}\right)
\end{aligned}
\end{equation}

In evaluating the integral in Eq. \ref{CAK} we use $\vec{n}_i = \sin\theta \vec{P} + \cos\theta \vec{N}$, where $\vec{N}$ is the unit vector pointing from the grid point towards the stellar center. $\vec{P}$ is a perpendicular vector chosen such that it is non zero and lies in a plane of the numerical grid. Integration is then performed numerically using a simple Simpson-rule with five steps in the interval $[0,\cos\theta_{\ast,i}]$.

Due to the ionization of the plasma, line driving is set to zero in cells with temperature above 10$^6$K.

\subsection{Radiative cooling}
Due to high densities and temperatures in the WCR, radiative cooling becomes important and thus has to be considered, see Eq. (\ref{euler3}). Here we use the cooling function $\Lambda(T)$ from \cite{Schure2009} who present a new radiative cooling curve based on an contemporary plasma emission code providing detailed logT-log$\Lambda$ tables. We use the tabulated data and interpolate between individual data-points.

\subsection{Geometrical setup \& initial conditions}
\label{geom}
We use a Euclidean coordinate system defined such that the x-axis coincides with the semimajor axis of the binary system, thus connecting periastron and apastron. The origin is at the center of mass. The z-axis is perpendicular to the orbital plane. 

Initially we place the stars along the x-axis (choosing between periastron and apastron passage) and initialize the stellar winds by a $\beta$-law approximation.

\begin{equation}
\label{mb1}
v_i(r)=v_{\infty,i} \left(1-\frac{\xi R_{\ast,i}}{r}\right)^{\beta_i}  
\end{equation}
From the continuity equation Eq. (\ref{euler1}) it follows that
\begin{equation}
\label{mb2}
 \rho_i(r)=\frac{\dot{M}_i}{4\pi r^2v_i(r)}
\end{equation}
for the individual stars. The temperature $T$ is initialized with the value 10$^4$K. It is not allowed to drop below this value, simulating the effects of photoionization heating. \citep[cf.][]{Pittard2009}.

As it is computationally expensive to self-consistently simulate the wind up from the stellar surface, it has been common practice to prescribe a fixed solution in at least 3 cells above the stellar surface at every time step \citep[e.g.][]{Pittard2009}. A classical $\beta$-law solution ($\xi=1$, only accurate in the point source limit) usually results in an non-physical kink between fixed cells and those dynamically solved. Multiple trials with 1D simulations clearly show that a smooth transition is best achieved by initiating the fixed cells with $\xi\sim0.9983$ -- which is also the value suggested by \cite{Cassinelli1999}. To achieve smoother wind-velocity profiles even at lower spatial resolutions we prescribe the wind with this modified $\beta$-law at every time step up to 5 cells above the stellar surface. For a typical resolution of $\sim$15.6 $R_\odot$ per cell, this gives satisfactory velocity and density profiles.   

For given stellar parameters, we determine the best fit value of $\beta$, as well as the values of the CAK-parameters $\alpha$ and $k$ by an iterative sequence of 1D simulations based on the known observables $R_\ast$, $L_\ast$, $M_\ast$, the wind's terminal velocity $v_\infty$ and the mass loss rate $\dot{M}$. 

To simulate the orbital motion of the stars, we use the standard equations for a Keplerian orbit. The differential equation for the eccentric anomaly $\Psi(t)$ (being $\omega t = \Psi - e \sin \Psi$ with eccentricity $e$) is  solved by a ten step Newton-Raphson method. 
The wind velocity that is prescribed at each step above the stellar surface is modified by the stellar velocity due to its orbital motion.

\section{PARTICLE SPECTRA} 
\subsection{The transport equation}
\label{transp}
In injecting electrons and protons inside the WCR which then suffer from various energy-loss mechanisms, we widely rely on the work by \cite{Reimer2006} in which an analytical approach for a simplified wind set-up is carried out in great detail.

We distinguish between \textit{acceleration cells} which are located at the shock front of the WCR, and all other cells. In order to discriminate between these two regions, we use the temperature structure of the wind plasma. Due to the imposed lower limit of 10$^{4}$ K, the wind is isothermal until it reaches the WCR. There, the temperature increases by three to four orders of magnitude within a few grid cells (see Fig. \ref{hydros}(b)). In our approach, we declare a cell to be an acceleration cell if three conditions are met: i) the divergence of the velocity vector be negative (the wind is effectively slowing down), ii) the temperature be higher than in the unshocked wind, and iii) the temperature of at least one of the six neighbouring cells be at the value of the unshocked wind. This yields a one-cell-thick skin of acceleration cells that envelop the WCR (see Fig. \ref{hydros}(d)).

The time-dependent transport equation - as it is solved for each grid point - reads:
\begin{equation}
\label{transport}
\frac{\partial N(E)}{\partial t}+\underset{(1)}{\nabla\cdot[\vec{v}N(E)]}
+\underset{(2)}{\frac{\partial}{\partial E}\left[\dot{E}N(E)\right]}
+\underset{(3)}{\frac{N(E)}{\tau}}=\underset{(4)}{Q_0\delta(E-E_0)}
\end{equation}
where $N$ is the differential number density of particles at energy $E$ in a grid cell at position $\vec{r}$.
We now discuss each term in detail:
\begin{enumerate} [label={(\arabic*)}]
\item The spatial convection term (where $\vec{v}$ is the velocity vector of the wind material) handles the transport of charged particles downstream along the WCR. It is used for all grid cells except acceleration cells which is treated in leaky box approach analogous to the acceleration region in \cite{Reimer2006}.
\item This term handles all energy gain and loss processes which depend on the considered particle species and on whether the cell is located at the shock front or not. For acceleration cells the term $\dot{E}$ includes DSA (not active outside the shock front) and radiative losses. We also consider adiabatic cooling which becomes important as the particles accelerate travelling downstream. For acceleration cells and all cells where $\nabla\cdot\vec{v}<0$ (true for cells just behind the shock front) the adiabatic cooling term has to be switched off as it would allow additional acceleration which is explicitly already taken care of by DSA.
\begin{equation}
\dot{E}=\left.
\begin{cases}
\dot{E}_\mathrm{DSA}+\dot{E}_\mathrm{radiative}\qquad &\mbox{for acceleration cells}\\
\dot{E}_\mathrm{radiative}\qquad &\mbox{elsewhere if $\nabla\cdot\vec{v}<0$}\\
\dot{E}_\mathrm{radiative}+\dot{E}_\mathrm{adiabatic}\qquad &\mbox{elsewhere if $\nabla\cdot\vec{v}>0$}\\
\end{cases}
\right.
\end{equation}
The individual energy gain and energy loss terms are discussed in detail in Sections \ref{accs} and \ref{lossmech}.
\item By considering the acceleration cells similarly to a leaky box model \citep[see e.g.][]{Protheroe1999}, the escape time $\tau$ describes the rate at which particles are lost by diffusing out of the system. 
The escape time can be approximated via the diffusion coefficient and the wind velocity  perpendicular to the shock $V_\mathrm{Shock}$ . For WCR cells outside the shock front, the diffusional leakage out of the system is set to zero by choosing an infinite escape time \citep[see][]{Martin2013}.
\begin{equation}
\tau=\left.
\begin{cases}
\frac{c_rD}{V_\mathrm{Shock}^2},\qquad\mbox{for acceleration cells}\\
\infty,\qquad\mbox{elsewhere}\\
\end{cases}
\right.
\end{equation}
with the compression ratio $c_r$ and the energy-independent diffusion coefficient $D$, which is an approximation for the sum of the upstream and downstream component of the diffusion coefficients at the shock ($D\approx D_1+c_rD_2$) \citep[e.g., ][]{Schure2010}.
\cite{Kirk1998} provide a comparison to the alternative assumption of an energy-dependent diffusion coefficient. In the present work, we choose the energy-independent approach in order to allow direct comparison with \cite{Reimer2006}. Thus, $D$ is constant throughout this work.
Note that an energy-independent diffusion coefficient demands additional consideration of the Bohm limit, at which further acceleration of particles is inhibited as their gyroradii become comparable to the characteristic size of the shock. We include this in our simulations by setting $\tau=0$ as soon as the Bohm diffusion coefficient ($\propto E$) exceeds the chosen energy-independent diffusion coefficient $D$.
\begin{equation}
\tau=0,\qquad\mbox{for } E>E_\mathrm{Bohm}=3eBD
\end{equation}
where the magnetic field strength $B$ is approximated as described below.

\item Particles are injected into the system at an energy $E_0$ (usually chosen to be 1 MeV). The choice of the injection rate $Q_0$  is limited to the constraints of particle number and energy conservation. A given grid cell in the shock region cannot inject more particles than it initially carries, neither can it deposit more energy in the injected particles than it has. Following \cite{Martin2013} we approximate the amount of injected particles by a constant fraction $\eta$ of the mass density.

Assuming a wind that predominantly consists of ionized hydrogen, partly ionized helium and electrons, it follows that
\begin{eqnarray}
n_e=\frac{\rho}{m_H}\frac{1+I_{He}\zeta_{He}}{1+4\zeta_{He}}\quad \mbox{and}\quad
n_p=\frac{\rho}{m_H(1+4\zeta_{He})}
\end{eqnarray}
with $n_p$ and $n_e$ being the number densities of free electrons and protons in a given cell, $\rho$ being the local wind density, $\zeta_{He}=\frac{n_{He}}{n_H}$, and $I_{He}$ being the number of electrons provided per helium nucleus. Hydrogen is assumed to be completely ionized. (We assume $\zeta_{He}=0.1$ and $I_{He}=2$ for wind temperatures higher than 30 000 K and $I_{He}=1$ below.)
This imposes a limit on the maximum number density for electrons $Q_0^e$ and protons $Q_0^p$ that can be accelerated per time step $d t$, of which we take the fractions $\eta_e$ and $\eta_p$. Other particle species liable for significant acceleration in the WCR (as He ions) are not considered in the present study.
\begin{equation}
Q_0=\left.
\begin{cases}
\frac{\eta_e}{dt}\frac{\rho}{m_H}\frac{1+I_{He}\zeta_{He}}{1+4\zeta_{He}}\quad \mbox{or}\quad
\frac{\eta_p}{dt}\frac{\rho}{m_H(1+4\zeta_{He})},\quad\mbox{for acceleration cells (electrons, protons)}\\
0,\qquad\mbox{elsewhere}\\
\end{cases}
\right.
\end{equation}
The injection fractions are used as free parameters. Typically, we take $\eta_p=10^{-3}$ and $\eta_e=10^{-5}$. The ensuing mass density decrease in the wind plasma is small enough to be negligible. As the energy deposited in the shock is also proportional to $\eta_{e,p}$ (and thus very small), there is no significant reduction of the kinetic energy of the wind. Energy conservation therefore holds.
We do not consider alternative sources of particles entering the acceleration process (e.g., via $\gamma-\gamma$ pair production.)
\end{enumerate}

\subsection{Particle acceleration}
\label{accs}
Of the various acceleration mechanisms for non-thermal particles that are discussed in literature, DSA (as a variant of the first-order Fermi acceleration principle in which particles gain energy by repeatedly traversing a shock) appears to be the most feasible process to be taken into account for our models. Alternative mechanisms as various turbulent processes (e.g., the second-order Fermi processes) or magnetic reconnection are assumed to be either less efficient or incapable of supplying particles that reach energies sufficient for  gamma-ray emission \citep[see][]{Dougherty2006}. 
Thus, we restrict ourselves to consider DSA as the sole acceleration process of electrons and protons, for the remainder of this work.
The average rate of momentum gain by DSA \citep[see e.g.;][]{Schure2010} is
\begin{equation}
\label{acc}
\dot{E}_\mathrm{DSA}=\left(\frac{c_r-1}{3c_r}\right)\frac{V_\mathrm{Shock}^2}{D}E
\end{equation}
where $V_\mathrm{Shock}$ is the shock velocity, $c_r$ the compression ratio and $D$ the energy-independent spatial diffusion coefficient as motivated above.

We define the shock velocity $V_\mathrm{Shock}$ as the upstream velocity normal to the shock and determine it by computing the velocity component of the plasma perpendicular to the WCR. As a tracer for the orientation of the collision region, we take the gradient of the temperature field $\vec{\nabla}T$ and compute
\begin{equation}
V_\mathrm{Shock} = \frac{\vec{v}\centerdot\vec{\nabla}T}{\mid\vec{\nabla}T\mid}
\end{equation}
The compression ratio $c_r$ at the shock is also directly determined from the wind structure by computing the ratio of post-shock mass density and pre-shock mass density. The former is determined by interpolating the mass density at a distance of three cell widths along the shock normal towards the WCR. The choice of three cell widths is motivated by the fact that the shock front in the simulations is in general three cells wide. A choice of two cell widths would yield too low postshock mass densities and result in a lower compression ratio. Likewise, a choice of four or more cell widths probes the mass density too far inside the WCR. As the mass density minimum of the wind plasma prior to reaching the WCR is generally located at the shock position for which the compression ratio is computed, we set the preshock mass density to the local value.

Once accelerated, the electrons and protons are subjected to several loss mechanism which we discuss below.

\subsection{Energy losses}
\label{lossmech}
\paragraph{\textbf{Inverse Compton emission} (electrons only)} 
IC cooling occurs as relativistic electrons scatter on stellar radiation fields within the binary system. It is a major energy-loss mechanism for non-thermal electrons. Here we use the full Klein-Nishina cross section resulting in the loss term:
\begin{equation}
\dot{E}_\mathrm{IC}=-b_\mathrm{IC} E^2f_\mathrm{KN}(E)
\end{equation}
with 
\begin{equation}
b_\mathrm{IC}=\frac{4}{3m_e^2c^3}\sigma_\mathrm{Th}u_\mathrm{ph}
\end{equation}
where $\sigma_\mathrm{Th}$ is the Thomson cross section and $f_\mathrm{KN}(E)$ is the correction factor for the Klein-Nishina regime \citep[following][]{Moderski2005} being
\begin{equation}
f_{KN}(\tilde{b})=\frac{9}{\tilde{b}^3}\left[\left(\frac{1}{2}\tilde{b}+6+\frac{6}{\tilde{b}}\right)\ln(1+\tilde{b})-\left(\frac{11}{12}\tilde{b}^3+6\tilde{b}^2+9\tilde{b}+4\right)\frac{1}{(1+\tilde{b})^2}-2+2\mathrm{Li}_2(-\tilde{b})\right]
\end{equation}
with the dilogarithm-function Li$_2$ and the dimensionless variable
$\tilde{b}=4\epsilon_T\frac{E}{m_e^2c^4}$
with $\epsilon_T$ being the energy of the target photon.
For $\tilde{b}<10^{-3}$ it is safe to set $f_\mathrm{KN}=1$.

The radiation energy density $u_\mathrm{ph}$ (taking into account both stars) is
\begin{equation}
u_\mathrm{ph}=\frac{1}{4\pi c}\left(\frac{L_{\ast,1}}{r_1^2}+\frac{L_{\ast,2}}{r_2^2}\right)
\end{equation}
where $L_{\ast,i}$ are the stellar luminosities and $r_i$ the distance to the stars.
At present, we do not consider additional radiation fields (e.g., photons from synchrotron emission) as targets for IC scattering.

\paragraph{\textbf{Synchrotron emission} (electrons only)}
The loss term for synchrotron emission is
\begin{equation}
\dot{E}_\mathrm{syn}=-b_\mathrm{syn} E^2
\end{equation}
with
\begin{equation}
b_\mathrm{syn}=\frac{4}{3m_e^2c^3}\sigma_\mathrm{Th}u_B
\end{equation}
where the magnetic energy density $u_B$ is defined as
\begin{equation}
u_B=\frac{B^2}{2\mu_0}
\end{equation}
Due to the lack of knowledge of the magnetic field strength $B$ in most colliding-wind binaries, we rely here on the approximations in \cite{Usov1992} who describe the magnetic field as either a classical dipole field, a radially dominated field and or a toroidally dominated field, depending on the distance from the star.
 \begin{equation}
B \approx\left.
\begin{cases}
B_\ast\left(\frac{R_\ast}{r}\right)^3, & \mbox{for } r<r_\mathrm{A} \\ 
B_\ast\left(\frac{R_\ast^3}{r_\mathrm{A} r^2}\right), & \mbox{for } r>r_\mathrm{A} \mbox{ and } r<R_\ast\frac{v_\infty}{v_\mathrm{rot}}\\
B_\ast\left(\frac{v_\mathrm{rot}R_\ast^2}{v_\infty r_\mathrm{A} r}\right), & \mbox{for } r>r_\mathrm{A} \mbox{ and } r > R_\ast\frac{v_\infty}{v_\mathrm{rot}} 
\end{cases}
\right.
\end{equation}
where $B_\ast$ is the magnetic field at the stellar surface and $v_\mathrm{rot}$ is the surface rotation velocity of the star, typically approximated by $v_\mathrm{rot}\sim0.1v_\infty$. For the Alfv\'{e}n radius $r_A$ we take\\
\begin{equation}
r_\mathrm{A} \approx\left.
\begin{cases}
R_\ast(1+\xi), & \mbox{for } \xi\ll1 \\
R_\ast\xi^{1/4}, & \mbox{for } \xi \gg 1 
\end{cases}
\right.
\end{equation}
with $\xi=\frac{B_\ast^2R_\ast^2}{\dot{M}v_\infty}$ .
As a reasonable value for the surface magnetic field $B_\ast=0.01$ T is assumed throughout this work for both stars \citep[see e.g.,][]{Reimer2006}.

Thus, we have an estimate for the magnetic field depending on the distance from each star. Since we merely approximate the absolute value of the magnetic field vector without having knowledge of its individual components, we cannot compute its vector sum for both stars. As an approximation, we merely consider the dominant component at a given grid cell to calculate $u_B$ and $b_\mathrm{syn}$.

In order to preserve consistency with \cite{Reimer2006}, we do not acknowledge the compression of the magnetic field in proportion to the compression of the wind inside the WCR. Thus, the field strength $B$ is likely to be underestimated by a factor of $\sim$4 which has no effect at lower energies, but can influence the maximum particle energy due to higher synchrotron losses and a higher Bohm-energy. In future work, a full MHD description of the wind will allow greater precision in the treatment of the magnetic field.

\paragraph{\textbf{Losses by thermal bremsstrahlung} (electrons only)}
As charged particles interact with the Coulomb fields of ions in the wind plasma, bremsstrahlung emission occurs. To account for the ensuing energy loss we use  
\begin{equation}
\dot{E}_\mathrm{br}=-b_\mathrm{br} E
\end{equation}
with
\begin{equation}
b_\mathrm{br}=\frac{2}{\pi}\alpha\sigma_\mathrm{Th}cN_\mathrm{H}
\end{equation}
where $\alpha$ is the fine-structure constant and $N_\mathrm{H}$ is the number density of the thermal ions in the wind which we approximate from the mass density in a given grid cell -- dividing it by $m_\mathrm{H}(1+4\zeta_{He})$.
\begin{equation}
N_\mathrm{H}\approx\frac{\rho}{m_\mathrm{H}(1+4\zeta_{He})}
\end{equation}

\paragraph{\textbf{Coulomb losses} (electrons)}
As loss term for the electrons we consider the energy-independent Coulomb losses given by
\begin{equation}
\dot{E}_\mathrm{coul}= -b_\mathrm{coul}=-55.725c\sigma_\mathrm{Th}N_\mathrm{H}m_ec^2
\end{equation}

\paragraph{\textbf{Coulomb losses} (protons)}
In the dense winds Coulomb losses can also become significant for protons. They are expressed by
\begin{equation}
\dot{E}_\mathrm{coul}=-\frac{3c\sigma_\mathrm{Th}m_ec^2Z^2\ln\lambda}{2}N_\mathrm{H}\frac{\beta^2}{x_m^3+\beta^3}
\end{equation}
with
\begin{equation}
\beta=\frac{\sqrt{E(E+2m_pc^2)}}{E+m_pc^2}
\end{equation}
and
\begin{equation}
x_m=0.2\sqrt{\frac{T_e}{10^8 \mathrm{ K}}}
\end{equation}
The term $\ln\lambda$ is the Coulomb-logarithm which we set to have the value 20. The electron temperature is assumed to be $T_e\sim$10$^8$K. 

\paragraph{\textbf{Nucleon-nucleon interaction} (protons only)}
For  nucleon-nucleon interactions the energy loss rate is
\begin{equation}
\dot{E}_\mathrm{pp}=-b_\mathrm{pp}E
\end{equation}
with
\begin{equation}
b_\mathrm{pp}=1.3\times3cN_\mathrm{H}\sigma_\mathrm{pp}\frac{m_\pi}{m_p}
\end{equation}
valid above the threshold for pion production at $E_\mathrm{thr}\simeq0.28 $ GeV. The cross section for nucleon-nucleon collision is $\sigma_\mathrm{pp}=3\times10^{-26} \mathrm{cm}^2$ and the factor 1.3 takes into account the element abundance ratio between H and He (9:1) assumed throughout this work.

\paragraph{\textbf{Adiabatic cooling}} For all grid cells outside the shock front with $\nabla\cdot\vec{v}>0$ we include the adiabatic cooling term 
\begin{equation}
\dot{E}_\mathrm{adiab}=\frac{E}{3}\nabla\cdot\vec{v}.
\end{equation}
The term becomes obsolete for $\nabla\cdot\vec{v}<0$ as it would then yield additional acceleration that is already physically taken care of by the DSA term.

\subsection{Maximum energies}
The delicate balance of DSA and the above mentioned loss terms is decisive for the shape and maximum energy of the electron and (to a lesser degree) the proton spectra. The final expression of $\dot{E}$ entering the transport equation for all acceleration cells is of the form 
\begin{equation}
\dot{E}= \dot{E}_\mathrm{DSA}-(b_\mathrm{IC}f_\mathrm{KN}(E)+b_\mathrm{syn})E^2-b_\mathrm{br}-b_\mathrm{coul}
\end{equation}
for the electrons and
\begin{equation}
\dot{E}= \dot{E}_\mathrm{DSA}-b_\mathrm{pp}E-\frac{3c\sigma_\mathrm{Th}m_ec^2Z^2\ln\lambda}{2}N_\mathrm{H}\frac{\beta^2}{x_m^3+\beta^3}
\end{equation}
for protons. 

In the case of electrons the approximate $E^2$-dependence of IC and synchrotron losses leads to a cut-off in the GeV range for an average WR-OB binary system. The specific energy of the cut-off and the question of whether IC or synchrotron losses are mainly responsible for it, depend on the energy densities of radiation and magnetic field. For some configurations where radiative losses are small, we see a cut-off due to the Bohm diffusion limit (as described in Section \ref{transp}(3) ).

For the protons, neither Coulomb losses nor losses by nucleon-nucleon interaction suffice to produce a cut-off. Rather, it is the Bohm diffusion limit (and thus the gyro-radii overcoming the characteristic size of the shock, as described above) that produces a cut-off - usually at a few TeV for an average WR-OB system. At low energies, however, Coulomb losses can inhibit particle acceleration at its early stage and lead to notable spectral features, even for protons.

\subsection{Implementation}
Solving the transport equation (\ref{transport}) is handled directly within the framework of the  \textsc{Cronos} code, where a semi-Lagrangian solver \citep[following][]{Crouseilles2010} has been implemented as additional module applied at each grid-point and time-step. 
To fully incorporate the particle spectra into the code, additional scalar fields are created -- one for each energy bin and particle species. Instead of having merely 5 fields ($\rho$,$v_x$,$v_y$,$v_z$,$T$) we use 205 -- meaning that we add 100 logarithmically equally spaced energy bins $[E_i,E_{i+1}]$ with $E_i\in$[1 MeV, 10 TeV], (i=1,...,100) for both electrons and protons. 
By treating those fields as advected scalars (similar to $\rho$ in Eq. \ref{euler1}), \textsc{Cronos} intrinsically handles spatial convection (term (1) in Eq. (\ref{transport})) via its HD solver. 

The aforementioned semi-Lagrangian solver then takes care of the remaining equation
\begin{equation}
\label{semiLagrapart}
\frac{\partial N(E)}{\partial t}+\frac{\partial}{\partial E}\left[\dot{E}N(E)\right]+\frac{N}{\tau}=Q_0\delta(E-E_0).
\end{equation}
where the inhomogeneity  on the right hand side can be treated as a simple boundary condition at the lowest energy bin of the spectrum where $N(E_0)\equiv\frac{Q_0}{\dot{E}(E_0)}$.

Equation \ref{semiLagrapart} is solved at each step after the application of the HD-solver. It usually uses the same time step as the HD part of the solver but is also does sub-cycling in the case that $\dot{E}$ is very large and thus the convection velocity in momentum space becomes too large (i.e. the distribution is shifted by more than one energy bin). This is most relevant  when electrons leave the shock, $\dot{E}_\mathrm{DSA}$ becomes zero and the remaining losses are very high. 
However, we find that
sub-cycling is rarely needed as the HD time step is small enough. For the example we discuss in Section \ref{results}, the applied HD time step is typically of the order of $\sim$100 seconds. For the chosen stellar and stellar wind parameters, the spectra in the shock at the apex of WCR typically need  $\sim 10^5$ seconds to build up until they reach convergence. If the acceleration is switched off and just the loss terms remain, the width of the bins at high energies is (due to the logarithmic scaling) still large enough in order to prevent that the shift is larger than one bin per time step. Even if it were otherwise, the implicit solver would yield a sufficiently accurate result.

The term $\frac{N}{\tau}$ in Eq. \ref{semiLagrapart}  is considered for acceleration cells only. As these are treated similar to leaky box where spatial convection from one cell to the other has been turned off, additional care must be taken concerning the transport from one acceleration cell to the next in the direction of the inner wind collision region. We deal with this problem by assuming that those particles vanishing from a single acceleration cell due to the diffusion term $\frac{N}{\tau}$ are not lost from the system but enter the next cell downstream of the shock. Further along, convection quickly becomes the dominant process for particle transport.

For acceleration cells the index of the resulting spectrum is highly dependent on the compression ratio of the wind. It can be easily shown that Eq. \ref{semiLagrapart} (for $\dot{E}=\dot{E}_{DSA}$ and the definitions above) has a solution $N(E)\sim E^{-p}$ with $p=\frac{c_r+2}{c_r-1}$. For a typical strong shock of $c_r=4$ we obtain $p=2$. If there is not compression at all and $c_r=1$ (no shock) the index approaches infinity as the spectrum disappears.

\section{RESULTS} 
\label{results}
\subsection{Models investigated}
Now, we want to illustrate the capability of our code and show first results on how the distribution functions of high energy electrons and protons evolve during a CWB's orbit. To allow for direct comparison and consistency checks we choose a system identical to the one used in the parameter studies in \cite{Reimer2006}. Table \ref{params} lists stellar and stellar wind parameters of the studied CWB with a B star and a WR star.  

The parameter $\eta$ which approximately marks the relative position of the WCR in between the two stars is determined by
\begin{eqnarray}
\eta=\frac{\dot{M}_1v_{\infty ,1}}{\dot{M}_2v_{\infty ,2}}
\end{eqnarray}
where the indices are assigned to the stars such that $\eta<1$. 
The relative distance of the WCR to the stars is approximately given by
\begin{eqnarray}
x_1=\frac{\sqrt{\eta}}{1+\sqrt{\eta}}\quad\mbox{and}\quad
x_2=\frac{1}{1+\sqrt{\eta}}
\end{eqnarray}

For OB-WR systems, the WCR is generally much closer to the OB star. 
This is the case for the system we discuss here which has $\eta=0.1$. 
 
\begin{figure}[h]
	\setlength{\unitlength}{0.001\textwidth}
	\begin{subfigure}{500\unitlength}
		\begin{picture}(500,420)
			\put(0,0){\includegraphics[trim=2.9cm 1.4cm 2cm 1cm, clip=true, width=500\unitlength]{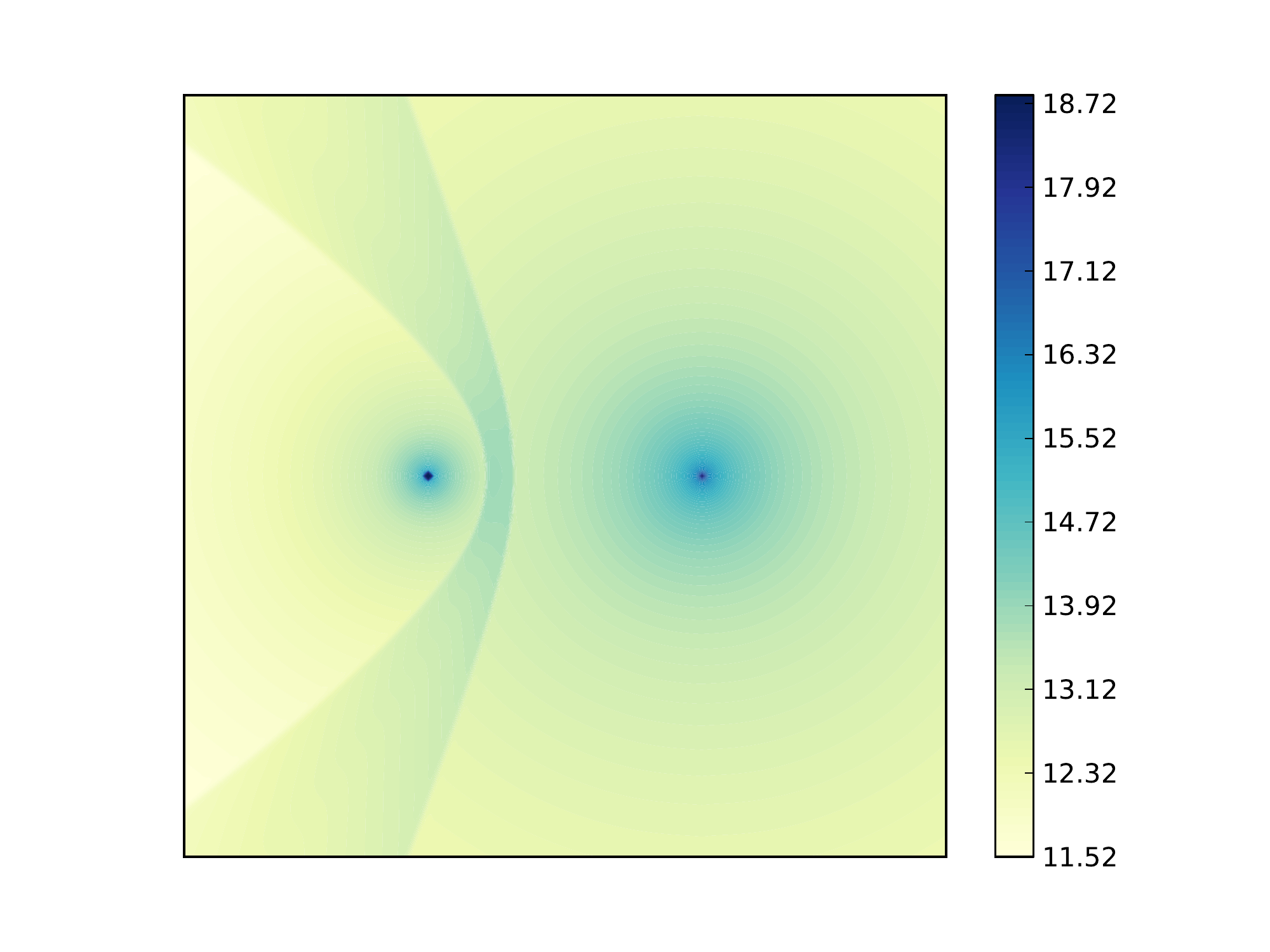}}
			\put(340,20){\Large{a})}
			\put(174.375,40){\line(1,0){49.25}}
			\put(174.375,36){\line(0,1){8}}
			\put(223.625,36){\line(0,1){8}}
			\put(174.375,15){\footnotesize{500R$_\odot$}}
			\put(400,150){\rotatebox{90}{\footnotesize{log(m$^{-3}$)}}}
		\end{picture}
	\end{subfigure}
	\begin{subfigure}{500\unitlength}
		\begin{picture}(500,420)
			\put(0,0){\includegraphics[trim=2.9cm 1.4cm 2cm 1cm, clip=true,width=500\unitlength]{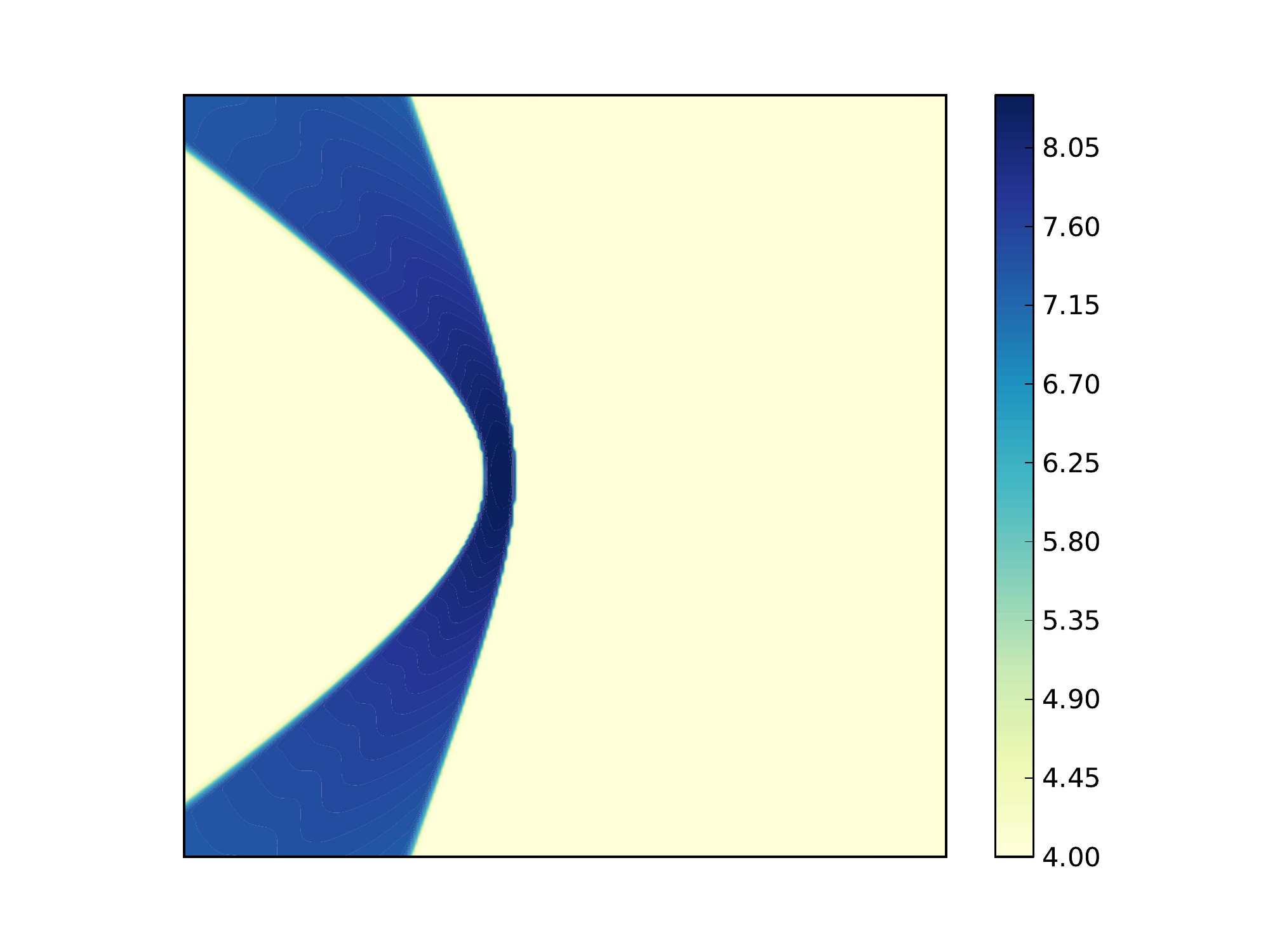}}
			\put(340,20){\Large{b})}
			\put(174.375,40){\line(1,0){49.25}}
			\put(174.375,36){\line(0,1){8}}
			\put(223.625,36){\line(0,1){8}}
			\put(174.375,15){\footnotesize{500R$_\odot$}}
			\put(400,150){\rotatebox{90}{\footnotesize{log(K)}}}
		\end{picture}
	\end{subfigure}\\
	\begin{subfigure}{500\unitlength}
		\begin{picture}(500,420)
			\put(0,0){\includegraphics[trim=2.9cm 1.4cm 2cm 1cm, clip=true, width=500\unitlength]{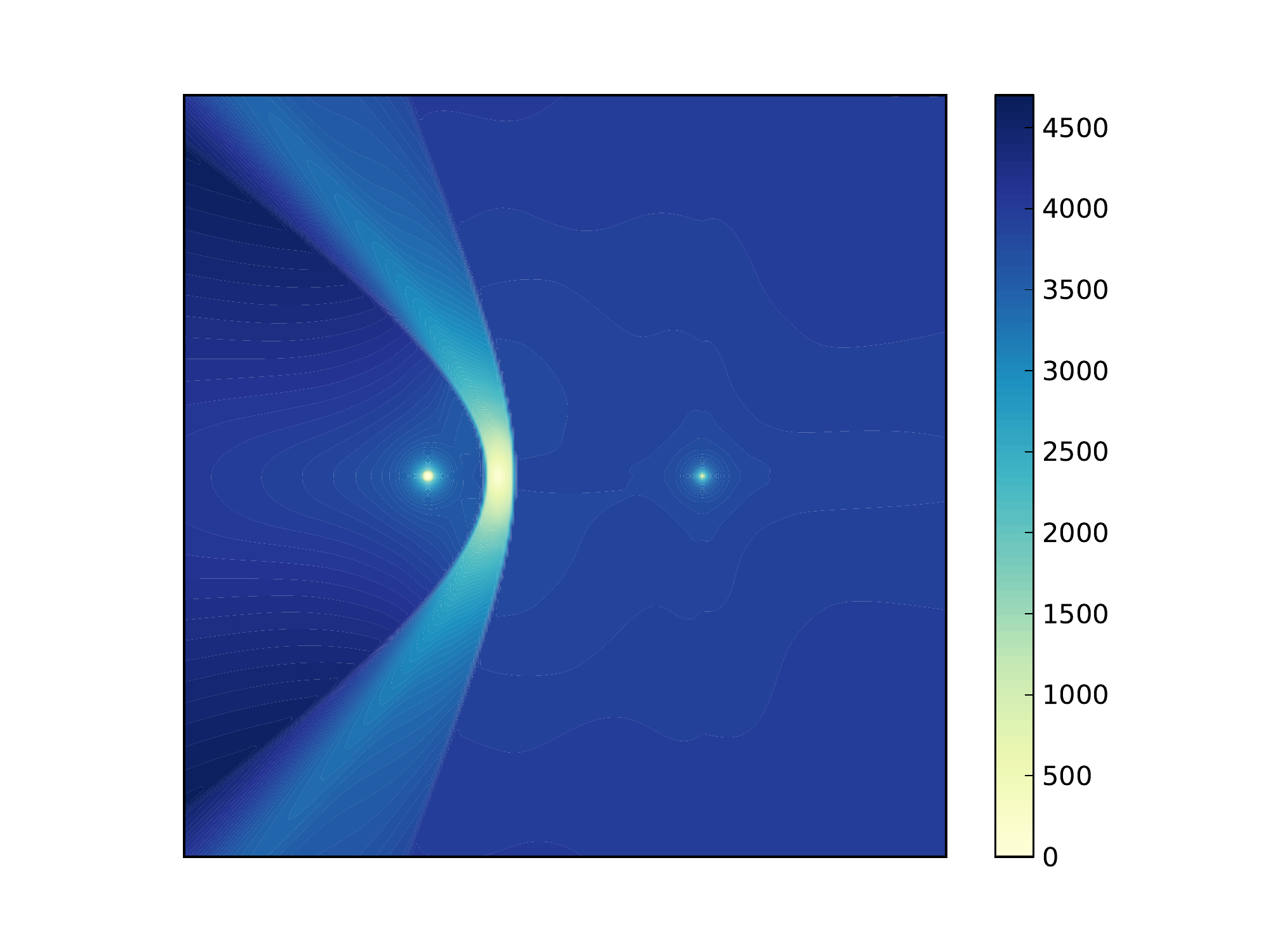}}
			\put(340,20){\Large{{\color{white}c)}}}
			\put(174.375,40){\color{white}\line(1,0){49.25}}
			\put(174.375,36){\color{white}\line(0,1){8}}
			\put(223.625,36){\color{white}\line(0,1){8}}
			\put(174.375,15){\footnotesize{\color{white}500R$_\odot$}}
			\put(400,150){\rotatebox{90}{\footnotesize{km s$^{-1}$}}}
		\end{picture}
	\end{subfigure}
	\begin{subfigure}{500\unitlength}
		\begin{picture}(500,420)
			\put(0,0){\includegraphics[trim=2.9cm 1.4cm 2cm 1cm, clip=true, clip=true,width=500\unitlength]{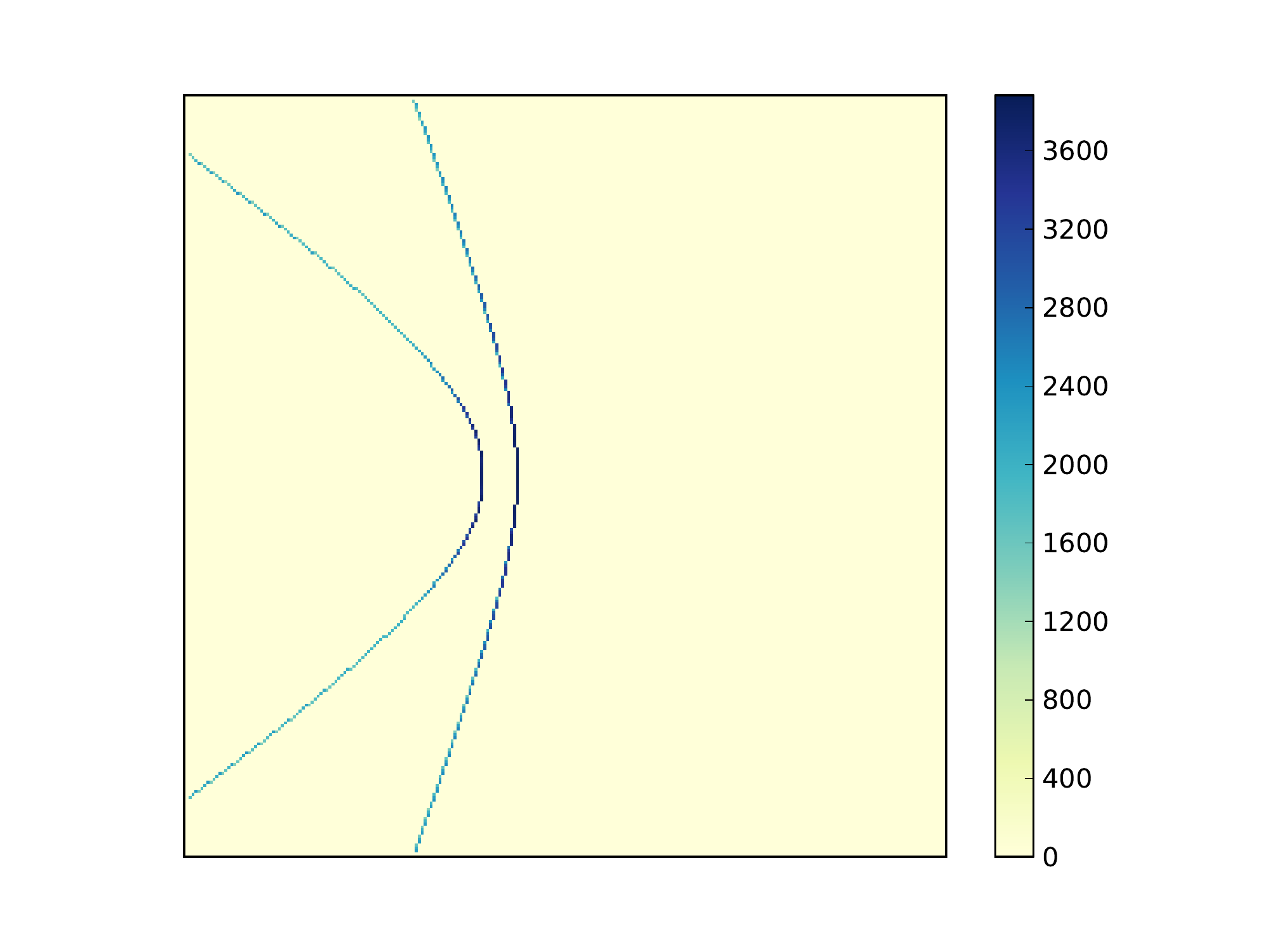}}
			\put(340,20){\Large{d})}
			\put(174.375,40){\line(1,0){49.25}}
			\put(174.375,36){\line(0,1){8}}
			\put(223.625,36){\line(0,1){8}}
			\put(174.375,15){\footnotesize{500R$_\odot$}}
			\put(400,150){\rotatebox{90}{\footnotesize{km s$^{-1}$}}}
		\end{picture}
	\end{subfigure}
\caption{HD quantities and shock front tracing in the converged state.   The plots show the x-y plane of a $256\times256\times64$ simulation  at z=0. 
The WR star is located on the right-hand side of the WCR, the B star on its left-hand side. The depicted quantities are a) particle density in log(m$^{-3}$), b) temperature in log(K), c) absolute velocity in km s$^{-1}$ and d) shock velocity in km s$^{-1}$.
}
\label{hydros}
\end{figure}

Fig. \ref{hydros}(a) shows the density structure obtained for the parameters given above. Without orbital motion the simulation converges quickly to the depicted state. Number densities of wind particles as high as $\sim10^{14}$ m$^{-3}$ are reached at the apex of the WCR. The WR star's higher mass-loss rate and higher wind momentum compared to its companion lead to a significant curvature of the WCR. Also note that the WCR is generally more dense on the side facing the WR star. 

\begin{table}[t]
\begin{center}
\textbf{Stellar and stellar-wind parameters}\\
\end{center}
\centering
\begin{tabular}{r ||c |c |c| l}
  & \textbf{B} & \textbf{WR}  & \textit{unit}\\ \hline \hline
 $M_\ast$ &30 & 30  & M$_{\odot}$\\
 $R_\ast$ &20 & 10 & R$_{\odot}$\\
 $T_\ast$ &23000 & 40000 & K\\
 $L_\ast$  & $10^{5}$ & $2.3\times 10^{5}$   & L$_{\odot}$\\
 $\dot{M}$ &$10^{-6}$ & $10^{-5}$  & M$_{\odot}$ yr\textsuperscript{-1}\\
 $v_\infty$ & 4000 & 4000  & km s$^{-1}$\\
 $\alpha$ & 0.643 & 0.678  & -\\
 $k$ & 0.5 & 0.7  & -\\
 $\beta$ & 0.83 & 0.83  & -\\
 \end{tabular}
\caption{Stellar and stellar-wind parameters of a typical binary system. The given CAK-parameters $\alpha$ and $k$ yield a stellar wind characterized by $\dot{M}$ and $v_\infty$ and a $\beta$-function type velocity law with the given $\beta$.\label{params}}
\end{table}

Fig. \ref{hydros}(b) shows the system's temperature distribution. As mentioned above the wind is kept at a constant temperature of $10^{4}$K outside the WR. It then heats up rapidly reaching temperatures $>10^{8}$ K at the apex. The figure suggests that temperature structure can be used as a tracer for the location of the shock fronts on the edge of the WCR.
	
Absolute wind velocity is shown in Fig. \ref{hydros}(a). The plot reveals how winds are slowed down to zero at the apex of WCR and then accelerate further downstream. Note that the regions where the wind of the B star reaches velocities exceeding its terminal velocity of 4000 km s$^{-1}$ can be explained by the combined radiative acceleration effect of both stars. It does not have any direct effect on WCR velocities as radiative acceleration is turned off for temperatures above 10$^{6}$K due to the fact that line driving does not affect a fully ionized plasma. However, the higher preshock velocities influence the postshock velocities which are highest at the wings (the outer part of the arms of the WCR) facing the B star's wind.

Fig. \ref{hydros}(d) shows the shock velocity which is the component of the wind velocity perpendicular to the shock. As it is undefined outside the shock, it also traces the acceleration region determined as outlined above. This is where DSA generates a population of relativistic particles which are then injected into the WCR where they advect downstream and lose energy.  

\subsection{Consistency checks}
\label{cons}
\begin{figure}
	\setlength{\unitlength}{0.001\textwidth}
	\begin{subfigure}[c]{500\unitlength}
		\begin{picture}(500,370)
			\put(0,0){
				\includegraphics[trim=0.2cm 0.2cm 0cm 0.0cm, clip=true,width=\textwidth]{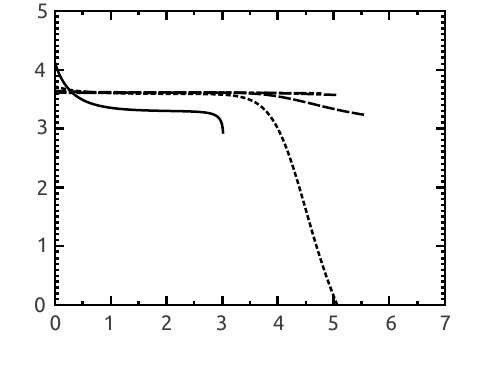}
			}
			\put(400,60){a)}
			\put(170,0){\footnotesize{log( E in MeV)}}
			\put(0,120){\rotatebox{90}{\footnotesize{log( $E^2N$ in MeV cm$^{-3}$ )}}}
		\end{picture}
	\end{subfigure}	
	\begin{subfigure}[c]{500\unitlength}
		\begin{picture}(500,370)
			\put(0,0){
				\includegraphics[trim=0.2cm 0.2cm 0cm 0.0cm, clip=true,width=\textwidth]{./f2a.pdf}
			}
			\put(400,60){b)}
			\put(170,0){\footnotesize{log( E in MeV )}}
			\put(0,120){\rotatebox{90}{\footnotesize{log( $E^2N$ in MeV cm$^{-3}$ )}}}
		\end{picture}
	\end{subfigure}	
\caption{a) differential number density of electrons for an acceleration cell after convergence. The five depicted cases (1 solid, 2 dotted, 3 dashed, 4 dash-dotted, and 5 double-dot-dashed) represent different values of  stellar separation $D$ $\approx$ 720, 1440, 2880, 7190, and 14380 R$_\odot$, magnetic field $B$ $\approx$ 1.3, 0.5, 0.3, 0.1, and 0.05 G, number density of the wind plasma $N_\mathrm{H} \approx$ (21, 5.2, 1.3, 0.2, and 0.05)$\times 10^{13}$m$^{-3}$ and shock velocity $V_\mathrm{s}\approx$ 1768,1884,1942,1977, and 1988 km s$^{-1}$.\newline
b) the same for protons }
\label{myels}
\end{figure}

Before applying the numerical solver for the transport equation to our HD simulations, we demonstrate consistency with previous studies: our treatment of the acceleration cells (see Section \ref{transp}) can be directly compared with the treatment of the acceleration region in \cite{Reimer2006}. For similar conditions in respect to shock velocity, density, magnetic and radiation field, our code reproduces spectra as shown in Fig. \ref{myels}.

One can identify the influence of various energy loss mechanisms. As the density of the wind plasma decreases steadily from case 1 to case 5, the effects of Coulomb losses  (visible for $E<$10 MeV) decrease accordingly (for electrons and protons). 
For electrons, the cut-off of the spectra at higher energies is either caused by synchrotron losses (cases 2 and 3), IC losses (cases 1 and 2) or by the diffusion approaching the Bohm limit (cases 4 and 5). 
The gentle slope visible for case 2 results from the curious condition that the IC loss rate almost reaches the acceleration rate at around 1~GeV, but fails to overcome it due the Klein-Nishina effect. The final cut-off at higher energies is due to synchrotron losses.
For protons the energy losses by nucleon-nucleon interaction and Coulomb losses do not suffice to overcome the acceleration. All the cut-offs are caused by the Bohm-limit.

We find  quantitative agreement of our results with \cite{Reimer2006} (i.e., Fig. 3 and 7). Minor differences of slope and cut-off shape in cases 1 and 2 can be understood by our usage of the full Klein-Nishina cross section. 

\subsection{Properties of accelerating cells}
\label{nuracc}
The same procedure as described above is now performed for all acceleration cells in our numerical grid. Shock velocity and number density of the plasma are directly determined from the HD variables. The magnetic and the radiation energy density are derived as outlined in Section \ref{lossmech}. At this point we are still neglecting the effects of orbital motion.

\begin{figure}
	\setlength{\unitlength}{0.001\textwidth}
	\begin{subfigure}[l]{500\unitlength}
		\begin{picture}(500,360)
			\put(15,23){
				\includegraphics[trim=0.7cm 0.9cm 0cm 0.1cm, clip=true,width=500\unitlength]{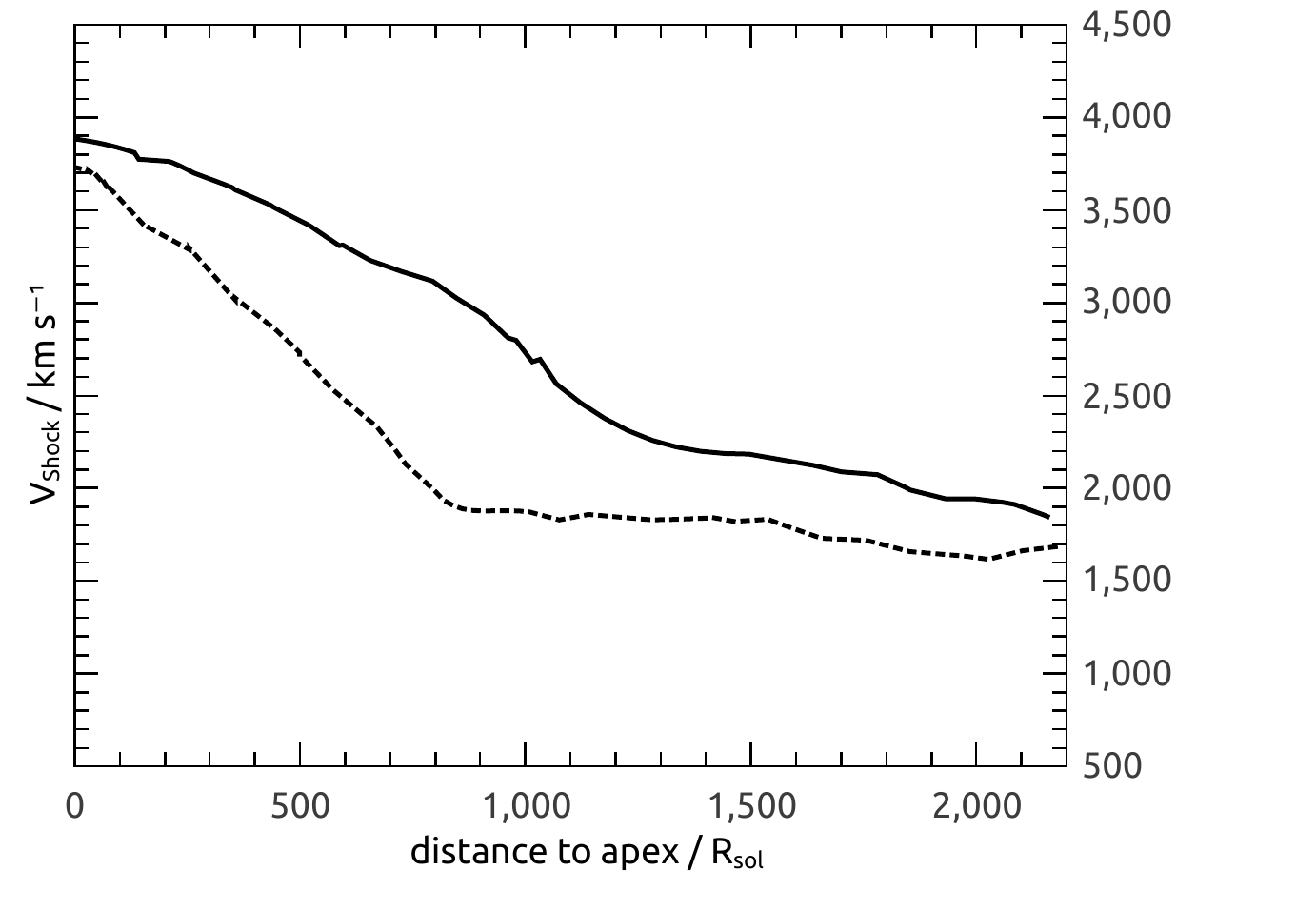}}
			\put(350,60){a)}
			\put(160,0){\footnotesize{distance to apex in R$_\odot$}}
			\put(0,120){\rotatebox{90}{\footnotesize{$V_\mathrm{Shock}$ in km s$^{-1}$}}}
		\end{picture}
	\end{subfigure}
	\begin{subfigure}[l]{500\unitlength}
		\begin{picture}(500,360)
			\put(15,20){
				\includegraphics[trim=0.7cm 0.9cm 0cm 0.1cm, clip=true,width=500\unitlength]{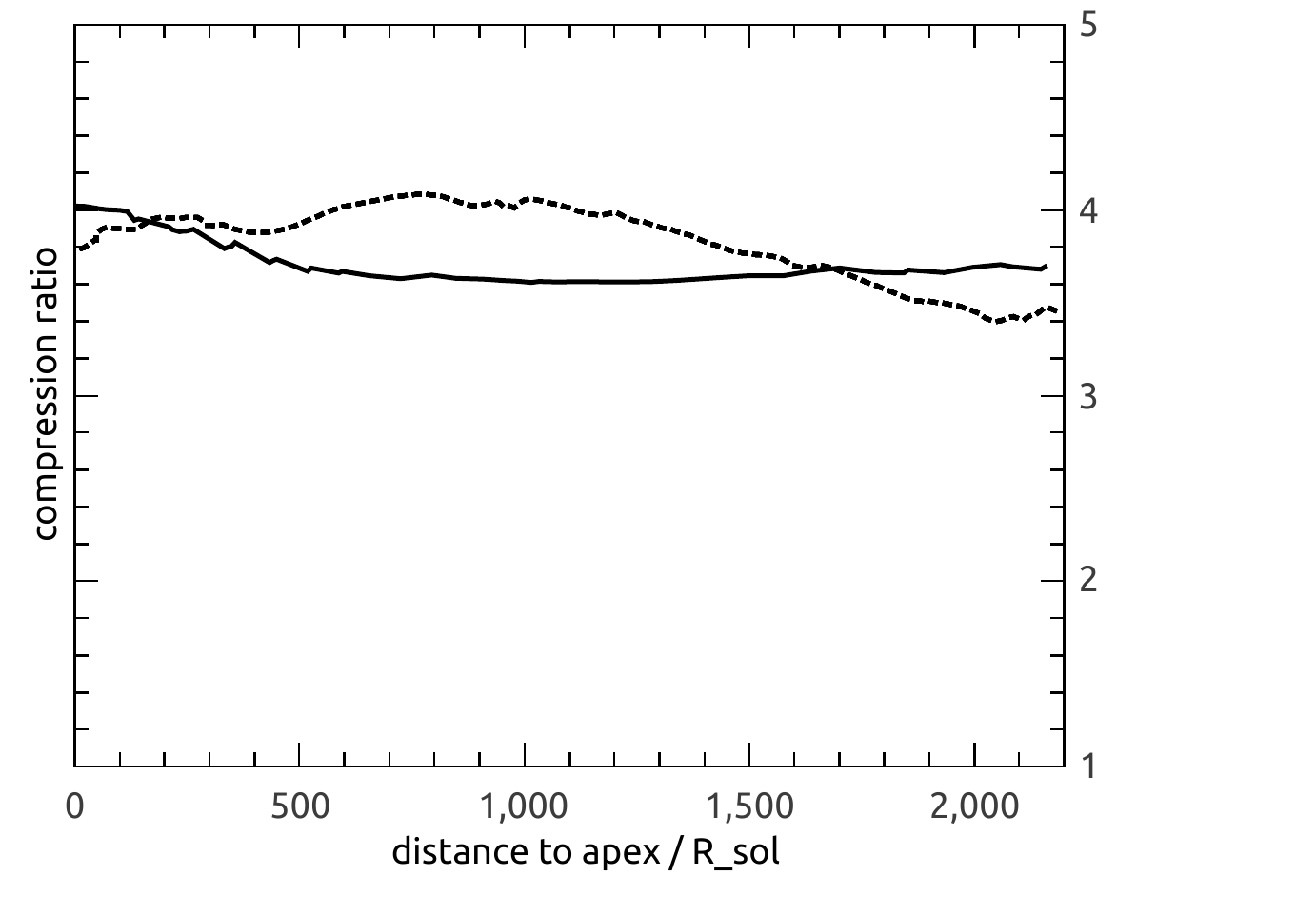}
			}
			\put(350,60){b)}
			\put(160,0){\footnotesize{distance to apex in R$_\odot$}}
			\put(0,90){\rotatebox{90}{\footnotesize{compression ratio}}}
		\end{picture}
	\end{subfigure}
	\begin{subfigure}[l]{500\unitlength}
		\begin{picture}(500,350)
			\put(15,22){
				\includegraphics[trim=0.7cm 0.9cm 0cm 0.1cm, clip=true,width=500\unitlength]{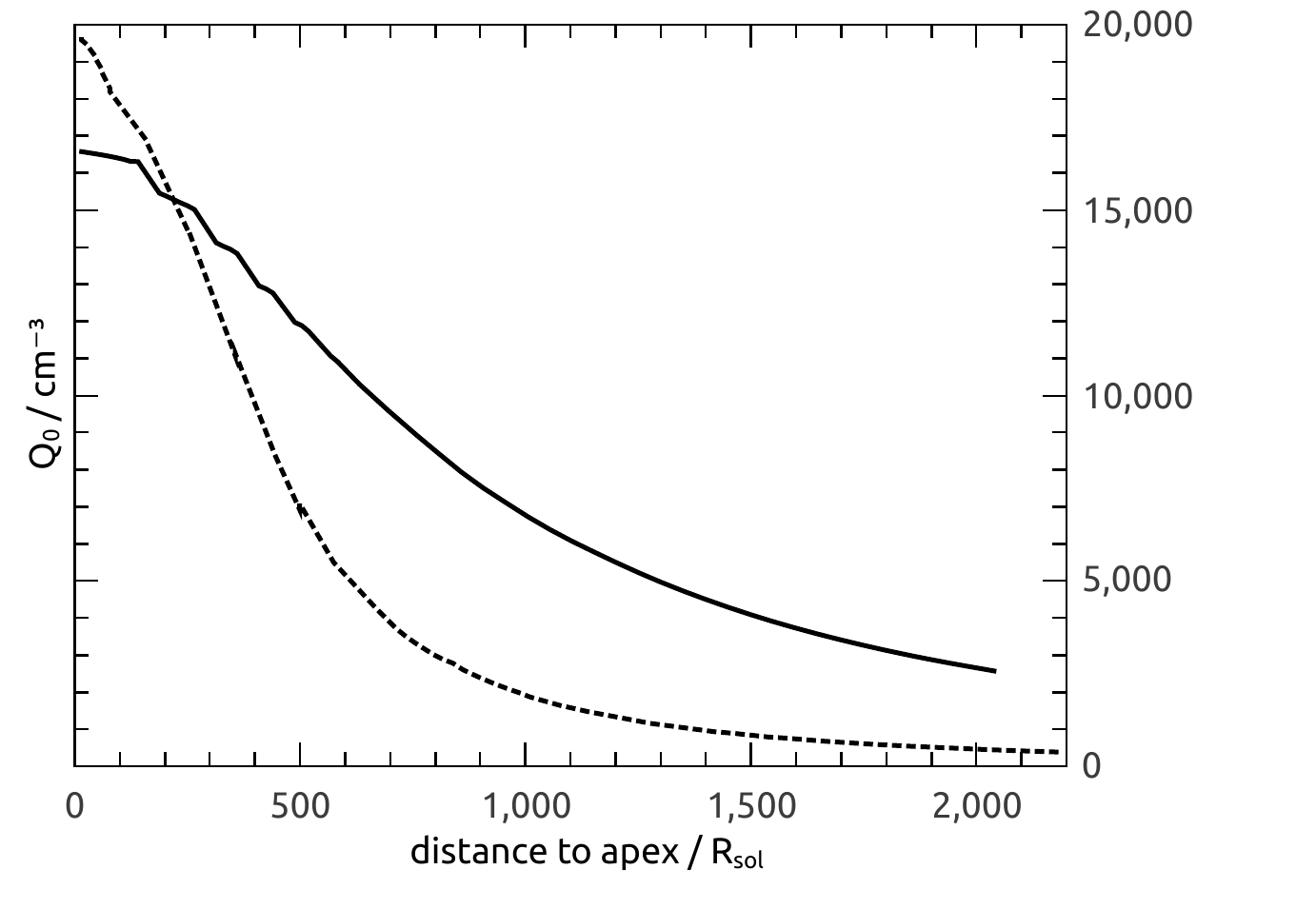}
			}	
			\put(350,60){c)}
			\put(160,0){\footnotesize{distance to apex in R$_\odot$}}
			\put(0,120){\rotatebox{90}{\footnotesize{$Q_0$ in cm$^{-3}$}}}
		\end{picture}
	\end{subfigure}
	\begin{subfigure}[l]{500\unitlength}
		\begin{picture}(500,350)
			\put(15,20){
				\includegraphics[trim=0.7cm 0.9cm 0cm 0.1cm, clip=true,width=500\unitlength]{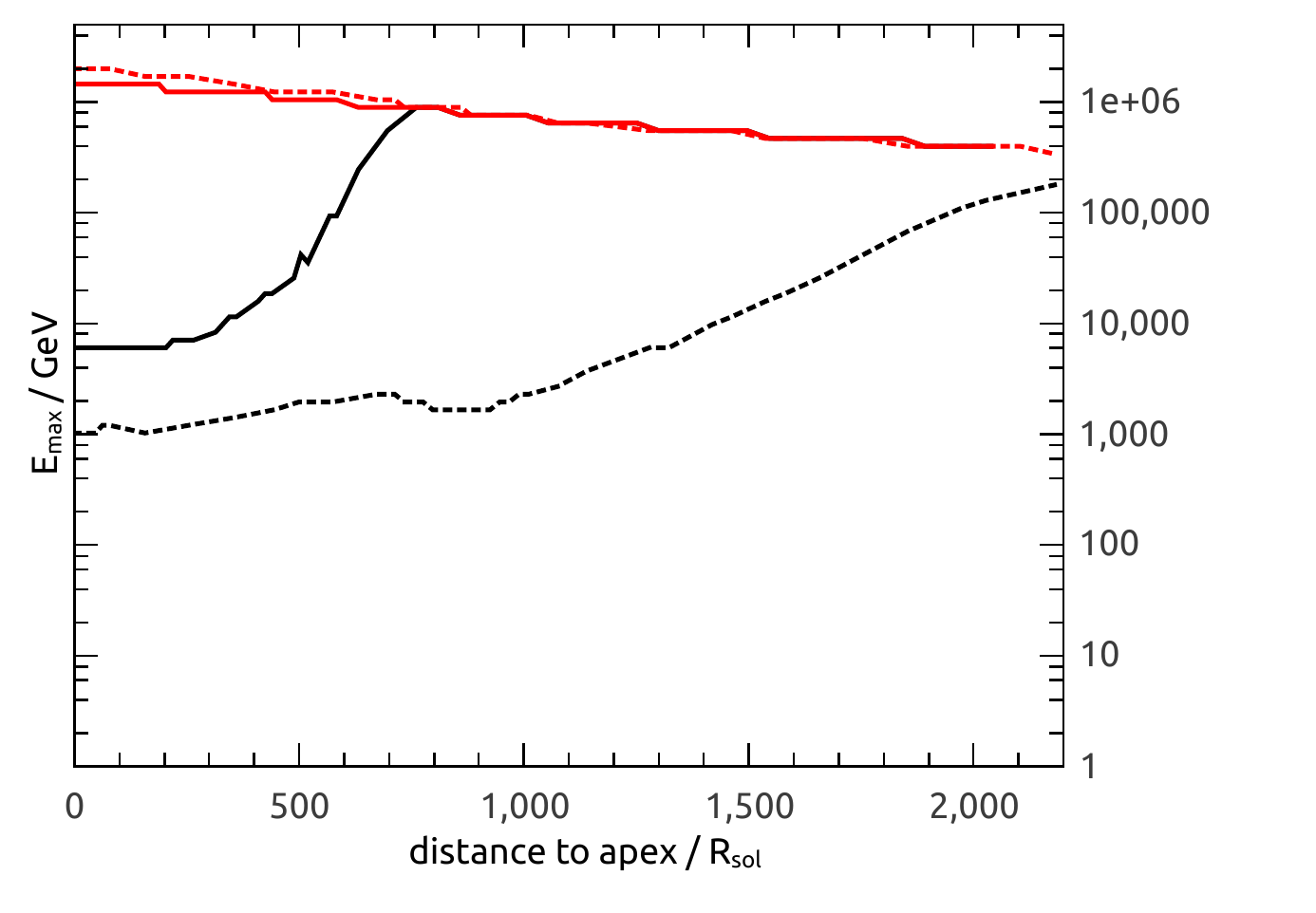}
			}
			\put(350,60){d)}
			\put(160,0){\footnotesize{distance to apex in R$_\odot$}}
			\put(0,120){\rotatebox{90}{\footnotesize{$E_\mathrm{max}$ in GeV}}}
		\end{picture}
			\end{subfigure}
			\caption{Important properties of the acceleration region along the shock as a function of the distance from the apex of the WCR: a) shock velocity, b) compression ratio (black), c) electron injection rate, d) maximum energy of electrons and protons. The solid lines represent the shock front facing the WR wind, the dotted lines represent the shock facing the B wind.
Concerning c), the proton injection rate is the same as for electrons multiplied by 
$\frac{\eta_p}{\eta_e(1+I_{He}\zeta_{He})}\approx100$.
In d) the maximum energies are shown for protons (red) and electrons (black).
					\label{props}}
\end{figure}
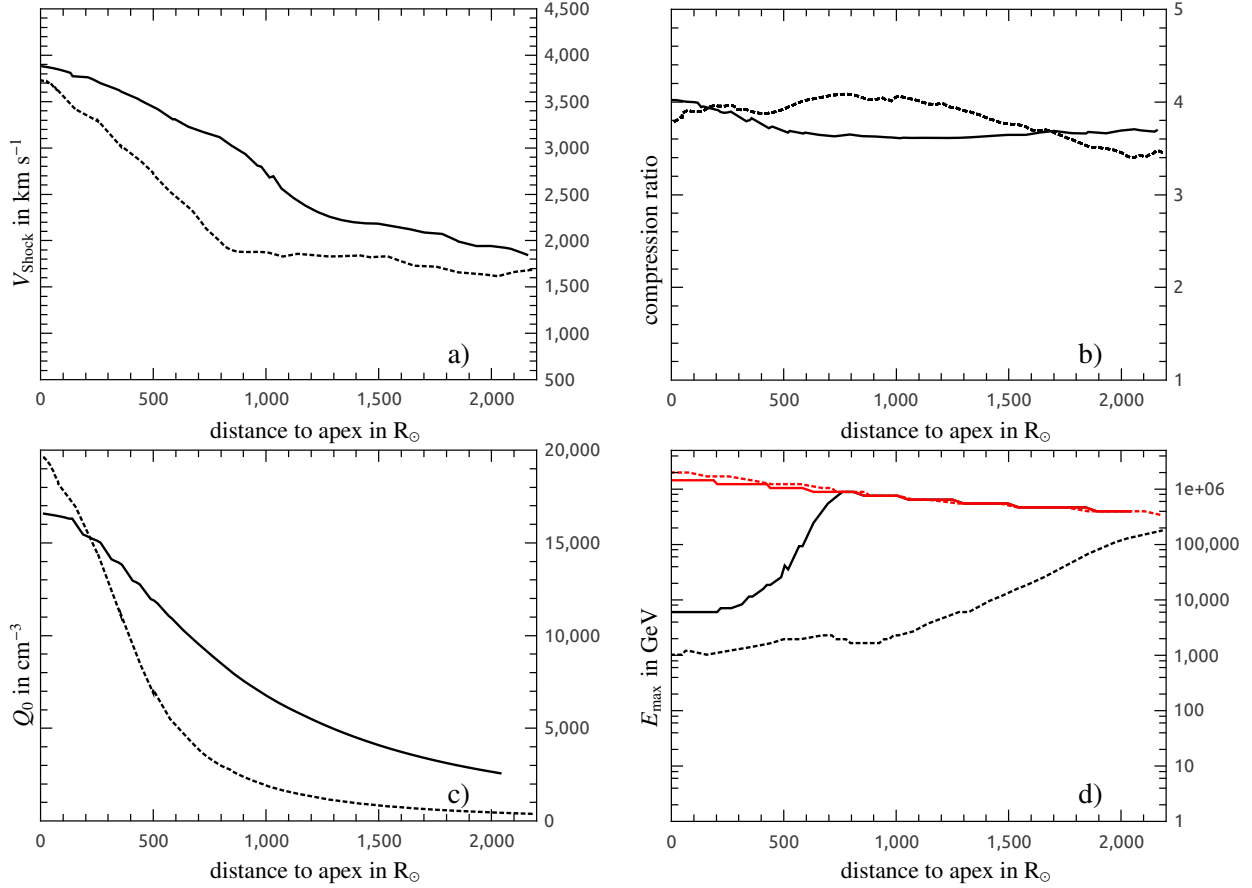

We explore the varying conditions along the shock front that determine how many particles are accelerated and how high their energies are. Fig. \ref{props} shows various properties that are significant for the acceleration process as they vary with growing distance from the apex along the thin surface of acceleration cells. A notable difference between the shock front facing the B star (in black) to the one facing the WR star (in red) becomes apparent. We will discuss each side in turn.

\paragraph{\textbf{The WR side of the shock}}
Looking at the values for the shock velocity $V_\mathrm{Shock}$ in Fig. \ref{props}(a) we see a slow and steady decrease with the distance from the apex. This decrease is caused by the shape of the WCR. As it curves around the B star, the velocity component along the shock decreases (even if the absolute wind velocities still increases). For distances from the apex greater than $\sim$1200 R$_\odot$ the curvature of the shock is less pronounced. Thus, the shock velocity decreases more slowly until it reaches zero at infinity.

Fig. \ref{props}(b) shows the compression ratio $c_r$. As expected for a strong shock it remains fairly constant, ranging from 3.8 to 4.
For compression ratios sufficiently above 1, the DSA-rate ($\dot{E}_{DSA}\sim\frac{c_r-1}{c_r}V_\mathrm{Shock}^2$) mainly resembles the behaviour of the shock velocity. The maximum is located at the apex. Acceleration then continuously decreases until far out in the wings.

As the injection rate of new particles is assumed to be proportional to the density in the acceleration cells, it decreases quickly with growing distance from the apex. This is shown in Fig. \ref{props}(c).

The maximum energy attained by the electrons depends on the various loss terms. IC scattering on photons in the radiation field of the B star is the dominant loss mechanism but it quickly looses relevance further away from the star. As the losses decrease faster than the DSA rate, Fig. \ref{props}(d) shows increasing maximum electron energies for increasing distances from the apex. This trend is broken at $\sim$800 R$_\odot$ when the electrons reach the Bohm-limit and, thus, cannot be accelerated any further. The slight decrease in maximum energy for even larger distances is due to the limit's proportionality to the magnetic field which is decreasing further out.
Also shown (in red) is the maximum energy of the protons which do not suffer any significant losses and attain their maximum energies close to the apex. 

\paragraph{\textbf{The B side of the shock}} 
As the curvature of the WCR is larger towards the B star's wind close to the star, the shock velocity drops more quickly than on the WR side. Also it is considerably smaller at the apex due to the lower terminal velocity of the B star's wind. As the shock front flattens out earlier than on the WR side, a lower rate of decrease in $V_\mathrm{Shock}$ is reached already at $\sim$800 R$_\odot$.

Again, the compression ratio remains fairly constant, ranging from 3.4 to 4.2. Minor fluctuations are expected to disappear with more sophisticated methods to determine the post-shock mass density and higher spatial resolutions. In general, the low level of variation in compression ratio along both shock fronts serves as confirmation that the generic assumption of fixing the $c_r$ to 4 is warranted for the kind of CWB system investigated.

Fig. \ref{props}(c) shows that the injection rate for the B side of the shock close to the apex is even larger than for the shock towards the high mass-loss WR wind. This is due to higher densities of the B wind in the proximity of the star. However, the steeper decrease in density soon overcomes the initially higher value. Further out, the injection ratio in the shock towards the B wind is up to an order of magnitude lower compared to the WR side of the shock. 

Fig. \ref{props}(d) reveals that the lower acceleration rate (due to lower $V_\mathrm{Shock}$ and $c_r$) and higher loss rates (due to the proximity of the B star) prevent the electrons on the B side of the shock to reach energies up to the Bohm limit until far out in the wings. The protons are again not affected by any significant losses. 

\subsection{Simulation results on particle spectra}
Having discussed the varying conditions along the shock, we now explore the resulting distribution functions of high-energy particles throughout the computational domain. We show particle distribution functions in the orbital plane for various energies as well as spectra for several selected positions.

\subsubsection{Electrons}
\begin{figure}
	\setlength{\unitlength}{0.001\textwidth}
	\begin{subfigure}[l]{900\unitlength}
		\begin{subfigure}{290\unitlength}
			\begin{picture}(290,290)
				\put(0,0){\includegraphics[trim=2.9cm 1.5cm 5.1cm 1.5cm, clip=true,width=290\unitlength]{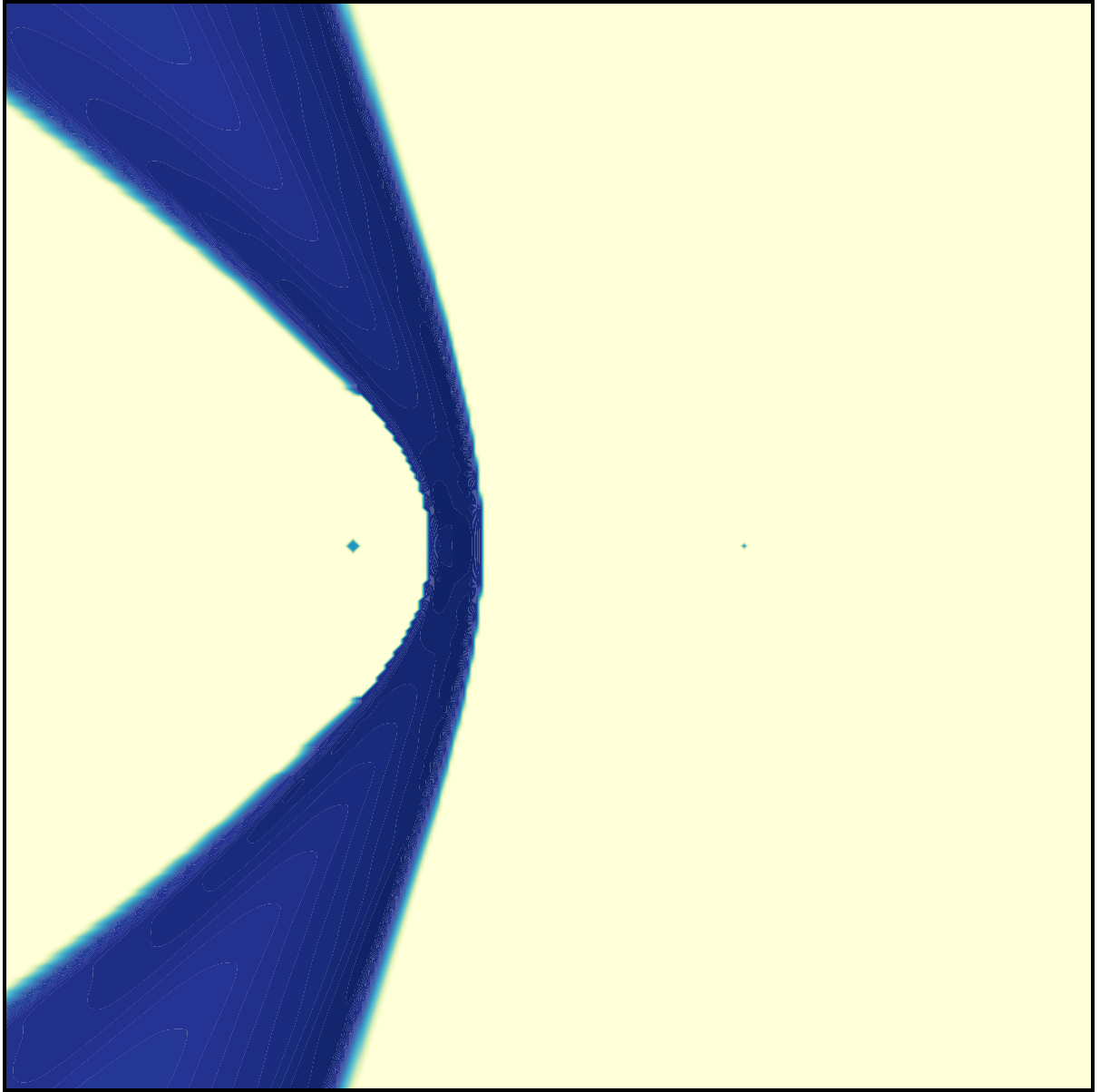}}
				\put(180,250){ 10 MeV}
				\put(126.7,20){\line(1,0){35.625}}
				\put(126.7,17){\line(0,1){6}}
				\put(162.325,17){\line(0,1){6}}
				\put(126.7,5){\scriptsize{500R$_\odot$}}
			\end{picture}
		\end{subfigure}
		\begin{subfigure}{290\unitlength}
			\begin{picture}(290,290)
				\put(0,0){\includegraphics[trim=2.9cm 1.5cm 5.1cm 1.5cm, clip=true,width=290\unitlength]{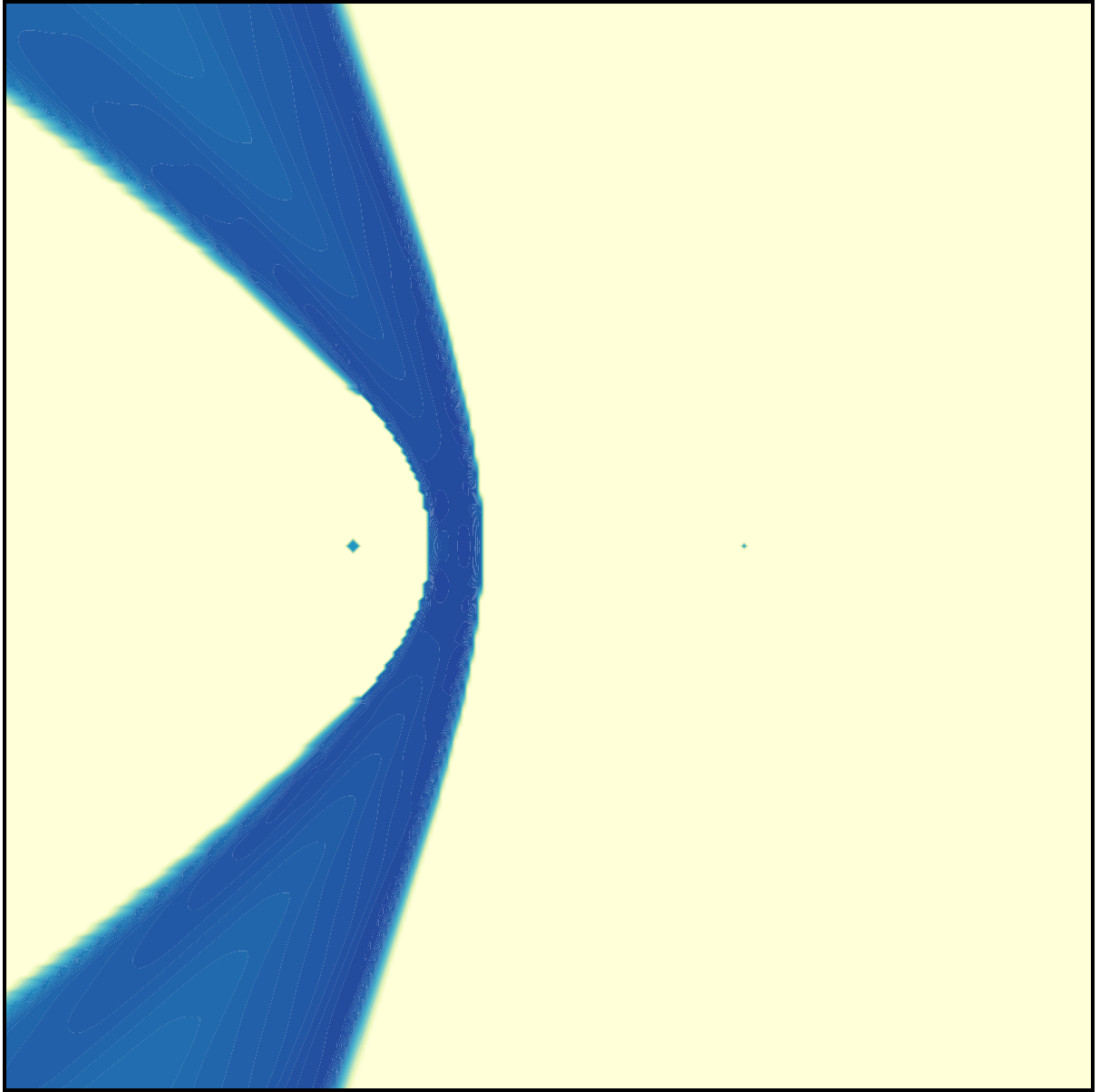}}
				\put(180,250){100 MeV}
				\put(126.7,20){\line(1,0){35.625}}
				\put(126.7,17){\line(0,1){6}}
				\put(162.325,17){\line(0,1){6}}
				\put(126.7,5){\scriptsize{500R$_\odot$}}
			\end{picture}
				\end{subfigure}
		\begin{subfigure}{290\unitlength}
			\begin{picture}(290,290)
				\put(0,0){\includegraphics[trim=2.9cm 1.5cm 5.1cm 1.5cm, clip=true,width=290\unitlength]{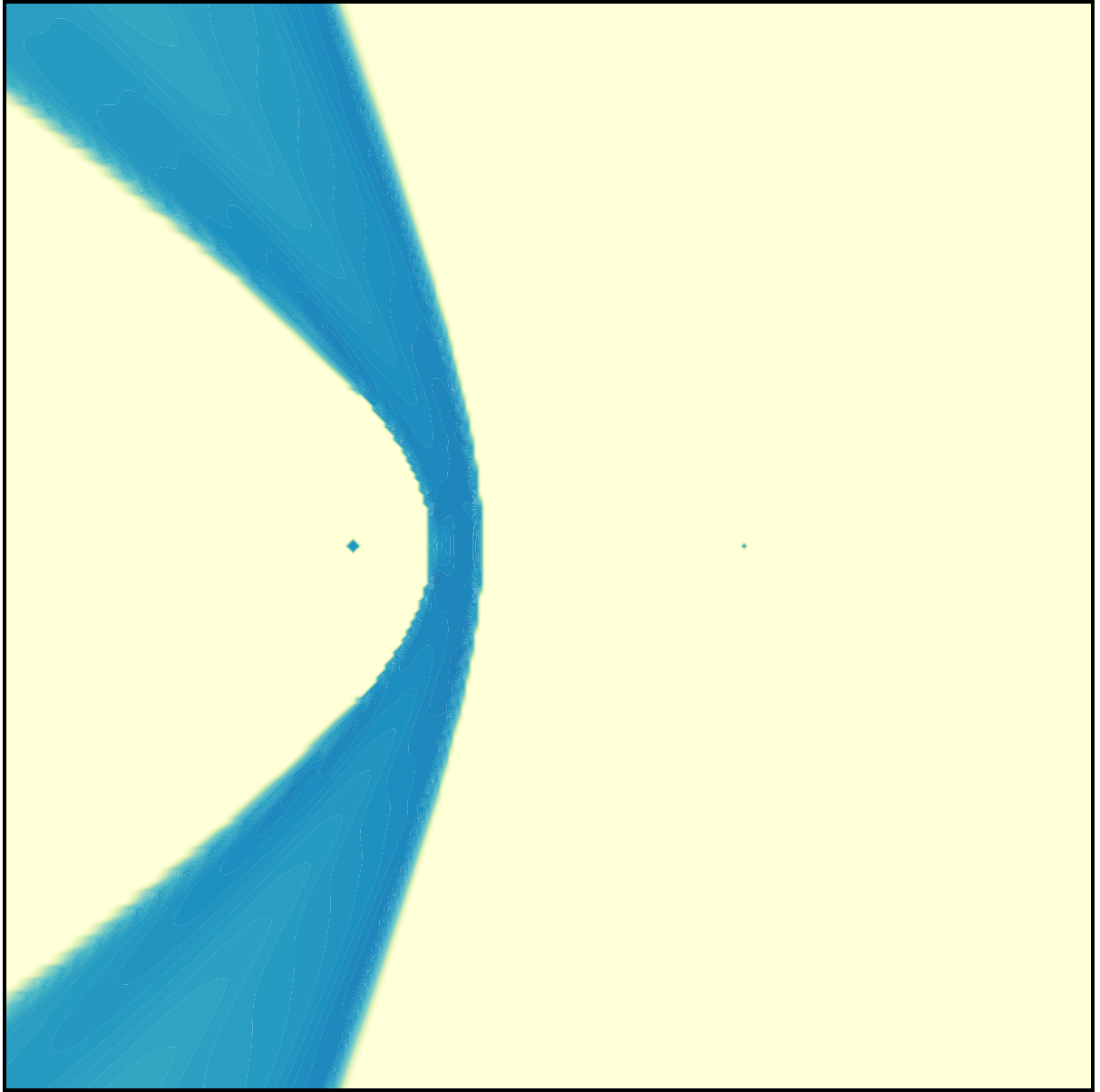}}
				\put(180,250){  1 GeV}
				\put(126.7,20){\line(1,0){35.625}}
				\put(126.7,17){\line(0,1){6}}
				\put(162.325,17){\line(0,1){6}}
				\put(126.7,5){\scriptsize{500R$_\odot$}}
			\end{picture}
		\end{subfigure}\\
		\begin{subfigure}{290\unitlength}
			\begin{picture}(290,290)
				\put(0,0){\includegraphics[trim=2.9cm 1.5cm 5.1cm 1.5cm, clip=true,width=290\unitlength]{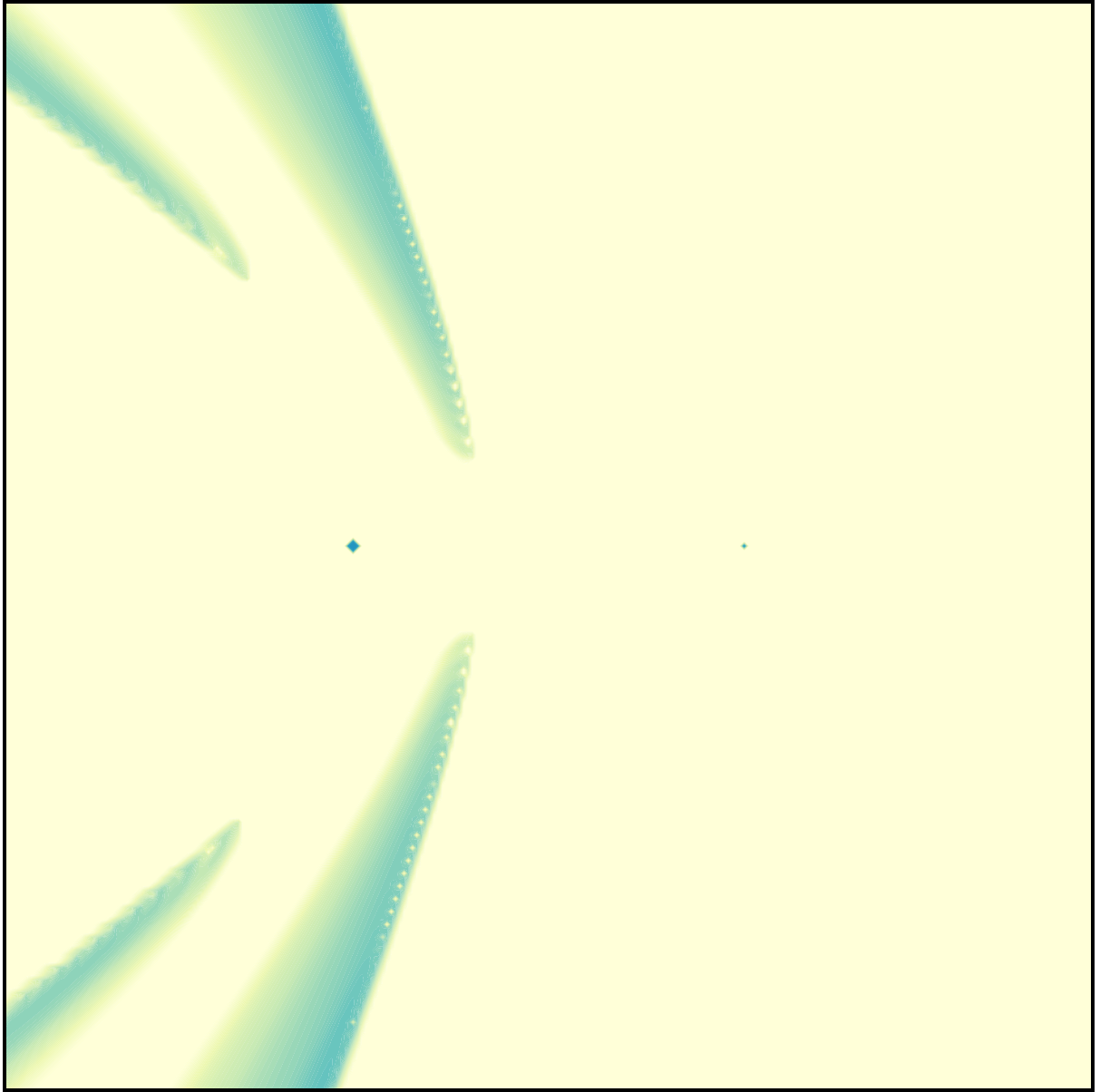}}
				\put(180,250){ 10 GeV}
				\put(126.7,20){\line(1,0){35.625}}
				\put(126.7,17){\line(0,1){6}}
				\put(162.325,17){\line(0,1){6}}
				\put(126.7,5){\scriptsize{500R$_\odot$}}
			\end{picture}
		\end{subfigure}
		\begin{subfigure}{290\unitlength}
			\begin{picture}(290,290)
				\put(0,0){\includegraphics[trim=2.9cm 1.5cm 5.1cm 1.5cm, clip=true,width=290\unitlength]{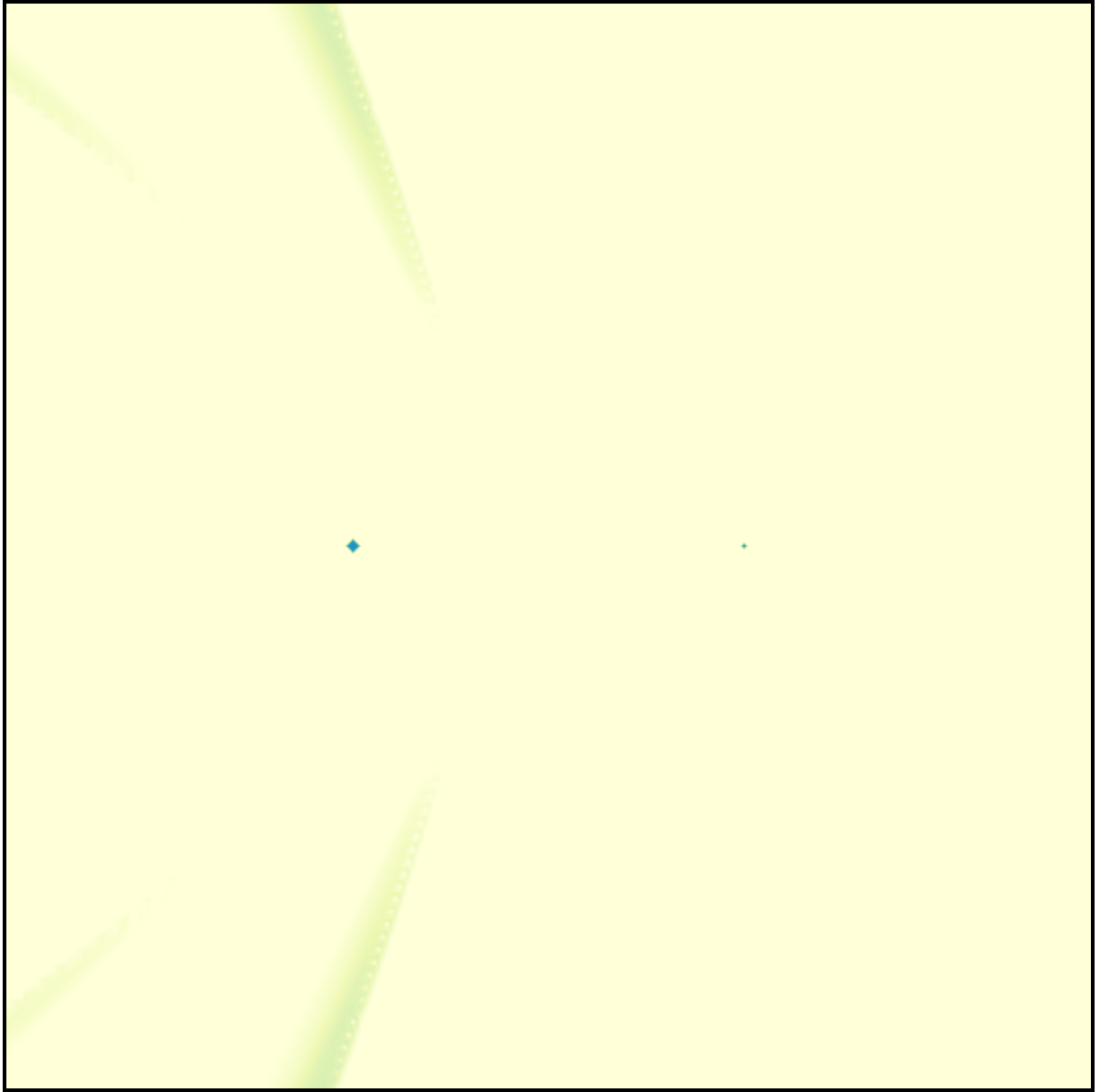}}
				\put(180,250){100 GeV}
				\put(126.7,20){\line(1,0){35.625}}
				\put(126.7,17){\line(0,1){6}}
				\put(162.325,17){\line(0,1){6}}
				\put(126.7,5){\scriptsize{500R$_\odot$}}
			\end{picture}
		\end{subfigure}
		\begin{subfigure}{290\unitlength}	
			\begin{picture}(290,290)
				\put(0,0){\includegraphics[trim=2.9cm 1.5cm 5.1cm 1.5cm, clip=true,width=290\unitlength]{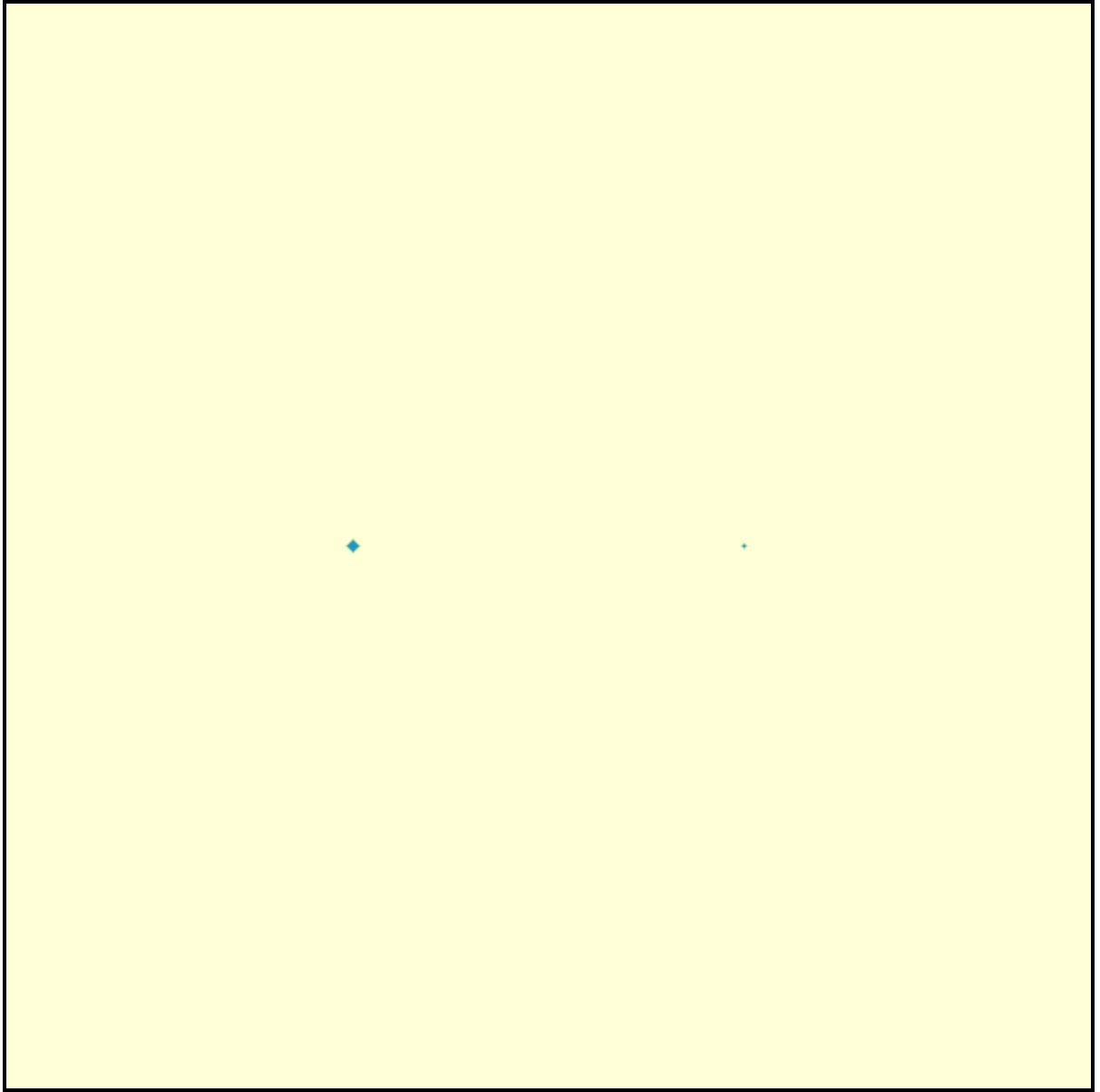}}
				\put(180,250){  1 TeV}
				\put(126.7,20){\line(1,0){35.625}}
				\put(126.7,17){\line(0,1){6}}
				\put(162.325,17){\line(0,1){6}}
				\put(126.7,5){\scriptsize{500R$_\odot$}}
			\end{picture}
		\end{subfigure}
	\end{subfigure}
	\begin{subfigure}{85\unitlength}
		\begin{picture}(85,580)
			\put(0,0){\includegraphics[trim=15.9cm 1.5cm 2.6cm 1.5cm, clip=true,height=580\unitlength]{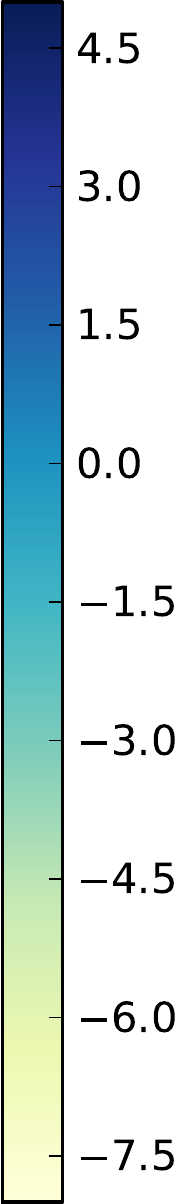}}
			\put(-24,200){\rotatebox{90}{log(MeV$^{-1}$m$^{-3}$)}}
		\end{picture}
	\end{subfigure}
\caption{Differential number density of electrons (in MeV$^{-1}$m$^{-3}$) as a function of kinetic particle energy. The colour maps show the x-y plane of a 256$\times$256$\times$64 simulation at z=0.   
\label{els}}
\end{figure}
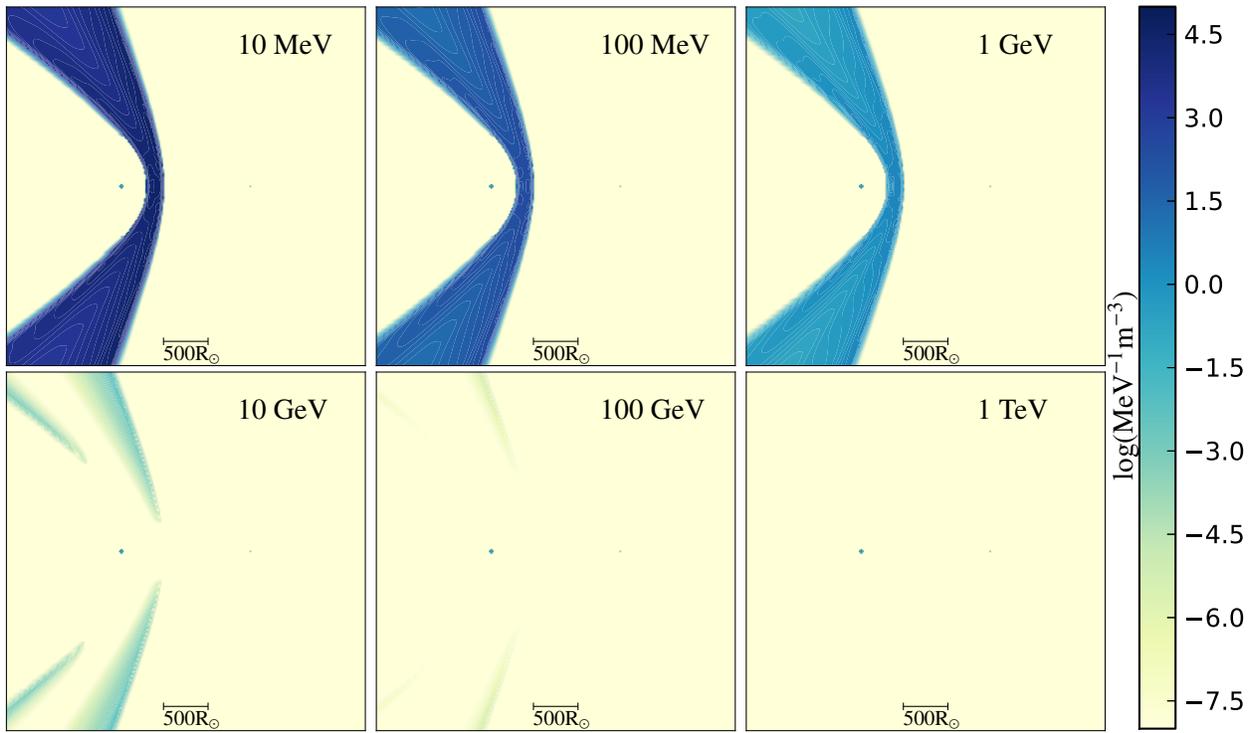

The electron distribution function for 6 different energies from 1 MeV to 1 TeV in the orbital plane is shown in Fig. \ref{els}. 

As expected, number densities drop quickly towards higher energies. It is interesting to observe that the highest energies are to be found in the wings of the WCR, close to the shock towards the WR star. This can be understood by considering that this is a region where radiative losses are low and the shock is strong, as the component of the wind velocity perpendicular to the WR side of the shock is still significant. Note that electrons at 10 GeV and higher can no longer penetrate into the center regions of the WCR due to severe radiation losses. They are confined to the proximity of the shock fronts. This behaviour closely resembles the schematic diagrams of the relativistic electron distribution in \cite{Pittard2006}.   

To allow a deeper understanding of the electron distribution function, Fig. \ref{elspec} shows a selection of spectra for different positions. 
Fig. \ref{elspec}(a) shows spectra along the apex of the WCR along the x-axis from right to left. The spectra at the shock fronts (solid and double-dot--dashed) are marked in red. As was shown above, the compression ratio at the WR side of the shock is close to 4 which yields a power law index of -2. Therefore, in E$^{2}$N scaling, we obtain a nearly flat spectrum with a cut-off where IC and synchrotron losses overcome the acceleration (red solid line).
Moving further into the WCR along the line connecting both stars we find higher electron densities than at the shock (black dashed line). 
This is caused by a pile-up of particles in the area behind the shock front which occurs due to very low fluid velocities in this region. As the spatial convection flux increases with number density, the cells behind the shock reach equilibrium between incoming and outgoing particles only at a certain number density which is generally higher than in the shock itself. 
Further to the center of the WCR (black dotted line), losses by cooling (which are larger at higher energies) become important. The spectra now shows two components, the one stemming from the B side of the shock with energies up to $\sim$ 1 GeV, the other stemming from the WR side with energies up to $\sim$ 10 GeV. Still closer to the B shock (black dash-dotted line) the latter component is further subdued as cooling and advection in both downstream directions remove more and more particles stemming from the WR side of the shock.  At the B shock (red double dot-dashed line), a lower compression ratio ($c_r\sim3.8$) yields a slightly softer spectral index $\sim2.1$. Radiative losses cause the change of slope and the eventual cut-off which is at a lower energy than on the other side of the WCR due to the increased proximity of the B star.

Fig. \ref{elspec}(b) shows spectra for several regions along the WR side of the shock front. In contrast to the spectra along the apex, it shows dramatic changes in maximum energy as loss rates decrease significantly with growing distance.  
Whereas it is still the IC losses that produce the cut-off close to the apex (solid and dashed line), the highest energies of the spectra further out are determined by the Bohm limit. However, the influence of the radiative losses can still be seen in the change of slope (dash-dotted and double-dot dashed line), very similar to the cases discussed in Section \ref{cons}.
A varying compression ratio 
leads to slight variations of the spectral index. Note that none of the spectra reaches energies $>$1 TeV. 

The spectra corresponding to the B side of the shock are shown in Fig. \ref{elspec}(c). 
We again see the expected increase in maximum energy as one moves towards the wings of the WCR. As radiative losses have a stronger impact on the B side of the shock, the Bohm limit is not reached except for a distance of $\sim$2000 R$_\odot$ from the apex. Additional features relate to slight variations in the compression ratio and to the decreasing injection rates as the wind plasma density decreases outwards.

Finally, Fig. \ref{elspec}(d) explores the spectra along the line that is equidistant to the two shock fronts. It shows a mixture of previously noticed effects. The absence of electrons above $\sim$1 GeV is due the fact that particles at those energies are produced at the shock front only at considerable distance from the apex. They do not have sufficient time to propagate into the center of the WCR before leaving the simulated region. Thus, the majority of the particles in this center region stem from close to the apex of the WCR.

\subsubsection{Protons}
Fig. \ref{prs} and \ref{prspec} are analogous to Fig. \ref{els} and \ref{elspec} for the case of protons. Radiative losses are hardly significant and thus, the number density of protons depends mostly on the injection rate (which is proportional to the local wind plasma density) than on the distance to the stars.
As radiative losses are of lesser importance, protons can reach energies of several TeV. Fig. \ref{prs} illustrates that, in contrast to electrons, the highest energy protons are to be found near the shock towards the B star where a stronger magnetic field strength shifts the Bohm-cut-off towards higher energies. 
Further details can be seen from the spectra in Fig. \ref{prspec} (a) to (d). Along the apex, we see again a pile-up behind the shock, as an equilibrium between the particle transport downstream by spatial convection and the flow of new particles from the shock by the diffusion term is not reached until high number densities are attained. Further away from the WR shock towards the center of the WCR, particles are increasingly transported downstream. The influence of the second shock front towards the B star emerges as the spectral index changes accordingly. There, a lower compression ratio yields a softer spectrum.

Comparing spectra along the both shock fronts (Fig. \ref{prspec}(b) and (c)), one can see various effects: the softening of the spectra with decreasing compression ratio, the decrease of the injection rate with decreasing mass density of the wind, and the slight decrease of the Bohm-limit with increasing distance from the apex. The latter explains why protons above 1 TeV are found near the center, but not in the wings of the WCR facing the WR star. Finally, Fig. \ref{prspec}(d) shows spectra along the center of the WCR which again are determined by the particle densities close to the apex. 

\subsection{Effects of orbital motion}
In addition to the previous findings discussed above, our simulations show that the orbital motion of a binary system has the potential to significantly alter the distribution of high-energy particles. As was described in Section \ref{geom}, our code is capable of moving the stars on a Keplerian orbit. In the present study, we merely investigate the circular case. 
By including orbital motion into our simulation and letting the system evolve for $\sim$1.3 orbital periods, we obtain results as in Fig. \ref{elsrot}. For the given stellar masses and their distance, a Keplerian orbit has a period of $\sim$820 days. The orbital velocity of the stars is $\sim$45 km s$^{-1}$ which is two orders of magnitude below the terminal velocity of the winds.

With orbital motion, the two arms of the WCR develop considerable differences. The effect becomes most notable at 10~GeV, where we see a contrast in differential number densities of $\sim$2 orders of magnitude between the leading and trailing arm on the B side of the WCR for electrons. This is due to the early cut-off of the trailing arm an energy of $\sim$10~GeV. Only in the leading arm, energies above 100 GeV are reached for the electrons along the shock towards the B star. For the WR side of the shock, maximum electron energies are reached in the trailing arm.

These differences stem to the greatest part from the deformation of the WCR due to orbital motion. It is no longer symmetric with respect to the apex. Fig. \ref{orbitspec}(a) illustrates the effect of these geometrical differences. The fraction of shock velocity versus absolute wind velocity is shown as a fraction of the distance from the apex along the shock front on the B side of the shock. The wind velocity component perpendicular to the shock is significantly larger along the forward arm. The ensuing difference in shock velocity has a significant impact as the acceleration rate is proportional to $V_\mathrm{Shock}^2$.  
Fig. \ref{orbitspec}(b) shows the electron spectra for two characteristic positions in the forward arm and in the trailing arm of the WCR close to the shock towards the B star. The lower shock velocity in the case of the latter causes a cut-off due to IC losses at $\sim$ 10 GeV. In case of the forward arm, the acceleration is able to compete with the losses up to higher energies. 

Comparatively larger effects are expected for elliptical orbits in which changing stellar separation causes more severe contrasts in a large number of relevant properties, as plasma density, radiation energy density, shock velocity, etc. In this work, we do not yet explore the effects of varying stellar separation. However, the significant differences between forward and trailing arm clearly indicate that orbital motion can have a strong influence in a model of the WCR and the particle transport within.

\section{DISCUSSION} 
We have demonstrated the feasibility of numerically solving the time-dependent transport equation of high-energy particles within a 3D hydrodynamic model of the stellar winds and their collision region for a massive star binary system.
This approach provides a more realistic description of the energetic particle spectra compared to analytical models with inherent simplifications regarding the complex structure of the WCR and the dynamics of the wind interaction.

For a typical CWB system of a WR and a B star where the WCR is much closer to the latter (e.g., $\eta$=0.1), we simulated the radiatively driven wind plasma and its 
collision region. 
On the basis of the HD description of the wind,the geometry of the WCR
and following the propagation of accelerating electrons and protons in the relativistic domain
, we derived relevant quantities related to particle acceleration such as shock velocity, compression ratio, maximum possible particle injection rate and maximum attainable energies along the two shock fronts. 
All of these exemplify notable differences between the B side and WR side of the shock and also between the inner (apex) and outer part (wings) of the WCR.
The geometrical structure of the WCR critically influences the shock velocity and, consequently, the maximum energy of electrons accelerated in the shock. 
Varying wind densities along the shocks affect the highest possible rate of particles injected at the shock.
We find a decrease in the shock velocity with growing distance from the apex that differs
corresponding to the dissimilar curvature and wind velocities at both sides of the WCR. By determining the compression ratio at the shock fronts directly from the mass densities provided by the HD simulation, we obtain values around $\sim$ 4. This corresponds well to the generic assumption of $c_r=4$ for strong shocks.  
Particle injection rates are largest close to the apex towards the B star due to its proximity. Further away from the apex, the denser WR wind produces a higher possible injection rate.

By solving a transport equation within the entire computational domain, we obtained spectra of high-energy particles throughout the WCR. For electrons, we have shown that maximum energies are attained in the wings of the WCR where losses by IC scattering and synchrotron emission are significantly lower than at the apex due to the distance from the stars. Typically, spectra with indices between 1.9 and 2.2 are obtained.
Since protons are less affected by energy loss mechanisms, their spectra show a cut-off determined by the Bohm limit. Varying indices and injection rates similarly relate to the wind properties and the geometry of the WCR. By studying spectra of high-energy particles in between the two shock fronts, we also provided insight into the evolution of particles after leaving the shock front on their journey downstream along the WCR. 

In addition, we demonstrated the importance of orbital motion which has considerable impact on the high-energy particle spectra. Even slight changes in the geometry of the WCR can cause contrasts in the number density of particles and critically influence their maximum energies.

\section{OUTLOOK} 
On the basis of the approach we presented, several applications and future developments become feasible. Concerning CWBs, the next consequent step is
to derive the photon emission along given line--of--sights on the basis of our electron and nucleon spectra.
The calculation of (anisotropic) IC, bremsstrahlung and neutral $\pi$-decay components of the total non-thermal photon flux can be done directly from the simulated particle population and will be presented in a subsequent paper. 
In addition, we will study the integrated flux value as a function of the stellar separation along the orbital period and as a function of time.
The application of this code to specific binary systems (e.g. WR 140) that remain undetected at $\gamma$-ray energies can provide limits on the fraction of injected electrons (which is an important free parameter in respective models). Dependencies of the $\gamma$-ray emission on other parameters, such as the magnetic field, can also be studied in detail. 
Future developments of the presented code include the consideration of additional physical processes, e.g. $\gamma$-$\gamma$ absorption close to the WCR.
The transition from a HD to a MHD description of the wind plasma holds promise to provide a self-consistent magnetic field model at the WCR that only depends on the magnetic field models of the stars.

The consideration of an increasing variety of aspects might eventually provide an explanation, why all CWBs except $\eta$~Carinae remain undetected at high-energy $\gamma$ rays so far. One critical factor that is now accessible to simulations is the dependence of particle evolution on the dynamics of the WCR, as it dramatically changes during the orbital period of high-eccentricity binary systems. Close to the periastron passage, especially strong cooling and strong velocity gradients produce very unstable conditions \citep[see e.g.,][]{Parkin2011,Madura2013} that could have a strong influence on the resulting particle distributions and non-thermal emission processes. This will be studied by a dedicated description of high-energy transport on the basis of 3D hydrodynamic models of colliding wind binaries.

\acknowledgments
 This work is supported through the EU FP7 programme by the Marie Curie IRG grant 248037, KR by the Marietta Blau-Stipendium der OeAD – GmbH, financed by the Austrian Ministry of Science BMWF. Support by the Austrian Ministry of Science BMWF as part of the UniInfrastrukturprogramm of the Research Platform Scientific Computing at the University of Innsbruck and the Austrian Science Fund (FWF) supported Doctorate School DK+ W1227-N16 is acknowledged.



\begin{figure}
	\setlength{\unitlength}{0.001\textwidth}
	\begin{subfigure}[c]{500\unitlength}
		\begin{picture}(500,370)
			\put(15,15){
				\includegraphics[trim=0.35cm 0.35cm 0.2cm 0cm,
				 clip=true,width=\textwidth]{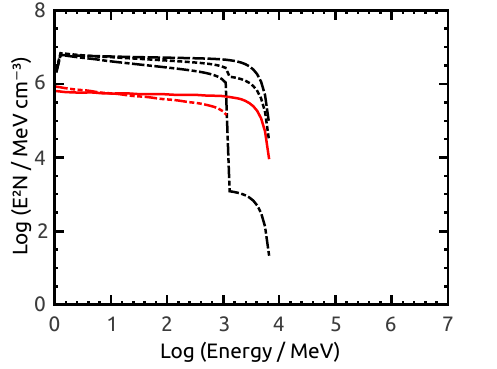}
			}
			\put(200,0){\footnotesize{log( E in MeV)}}
			\put(0,120){\rotatebox{90}{\footnotesize{log( $E^2N$ in MeV cm$^{-3}$ )}}}
			\put(440,80){a)}
		\end{picture}
	\end{subfigure}
	\begin{subfigure}[c]{500\unitlength}
		\begin{picture}(500,370)
			\put(15,15){
				\includegraphics[trim=0.35cm 0.35cm 0.2cm 0cm,
				 clip=true,width=\textwidth]{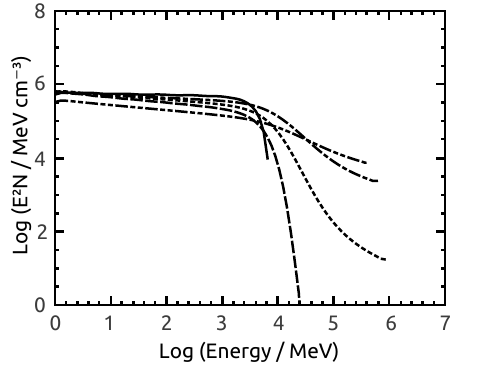}
			}
			\put(200,0){\footnotesize{log( E in MeV)}}
			\put(0,120){\rotatebox{90}{\footnotesize{log( $E^2N$ in MeV cm$^{-3}$ )}}}
			\put(440,80){b)}
		\end{picture}
	\end{subfigure}
	\begin{subfigure}[c]{500\unitlength}
		\begin{picture}(500,400)
			\put(15,15){
			\includegraphics[trim=0.35cm 0.35cm 0.2cm 0cm,
				 clip=true,width=\textwidth]{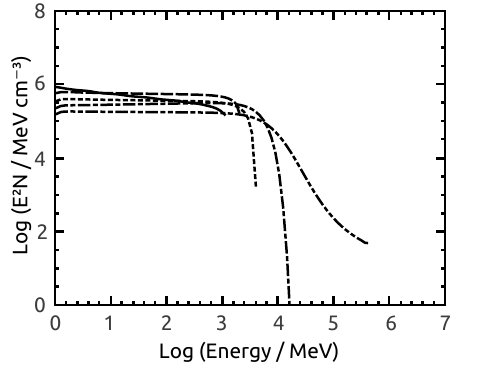}
			}
			\put(200,0){\footnotesize{log( E in MeV)}}
			\put(0,120){\rotatebox{90}{\footnotesize{log( $E^2N$ in MeV cm$^{-3}$ )}}}
			\put(440,80){c)}
		\end{picture}
	\end{subfigure}
	\begin{subfigure}[c]{500\unitlength}
		\begin{picture}(500,400)
			\put(15,15){
			\includegraphics[trim=0.35cm 0.35cm 0.2cm 0cm,
				 clip=true,width=\textwidth]{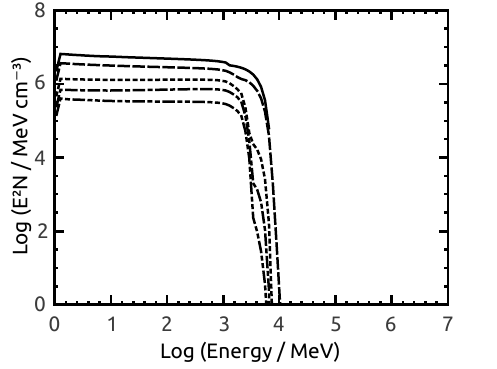}
				}
			\put(200,0){\footnotesize{log( E in MeV)}}
			\put(0,120){\rotatebox{90}{\footnotesize{log( $E^2N$ in MeV cm$^{-3}$ )}}}
			\put(440,80){d)}
		\end{picture}	
			\end{subfigure}
		\caption{Electron spectra for various positions within the WCR. Fig. (a) shows spectra along the connecting line of the stars, where distance from the shock front facing the WR wind (solid) is $\sim$60 (dashed), $\sim$110 (dotted), $\sim$160 (dash-dotted) and $\sim$190 R$_\odot$ (double-dot--dashed). Spectra within the acceleration region are coloured in red. Fig. (b) to (d) show regions along the WR shock (b), along the B shock (c) and along the centre of the WCR (d). Here, the distance to the corresponding region at the apex (solid) is  $\sim$500 (dashed), $\sim$1000 (dotted), $\sim$1500 (dash-dotted) and $\sim$2000 R$_\odot$ (double-dot--dashed).  \label{elspec} }
			
\end{figure}
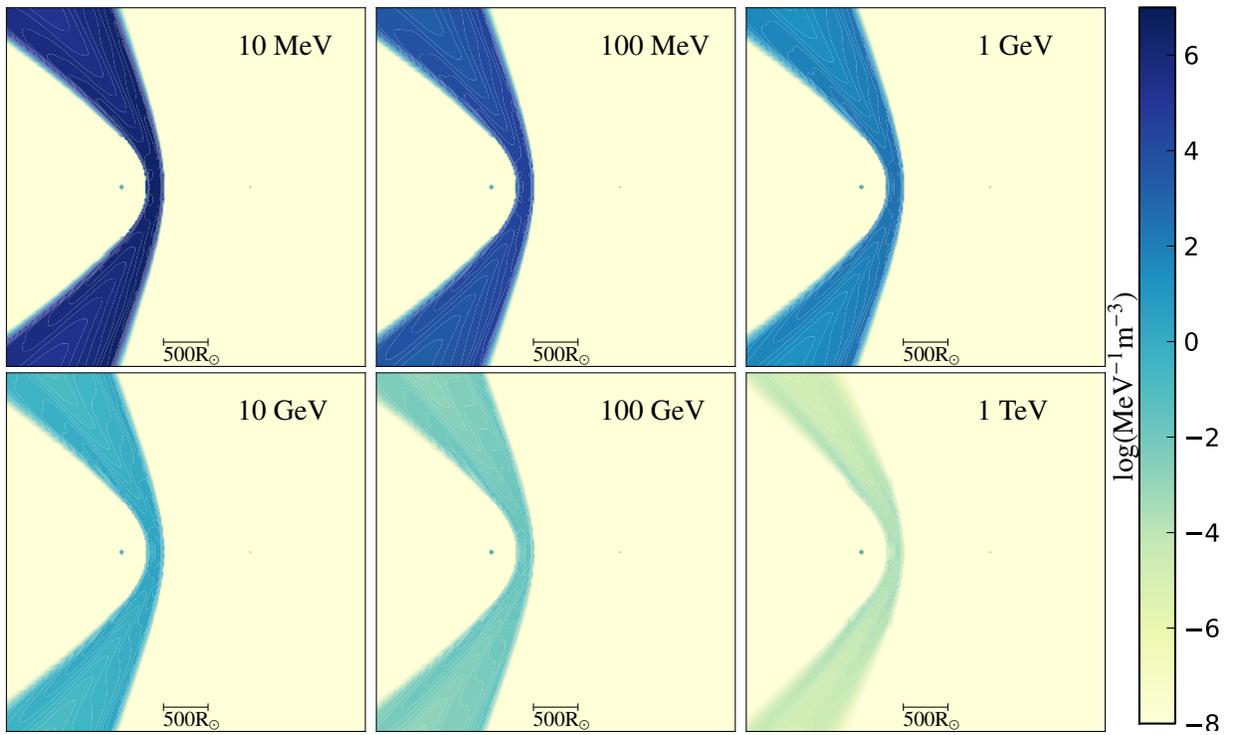
\begin{figure}
	\setlength{\unitlength}{0.001\textwidth}
	\begin{subfigure}[l]{900\unitlength}
		\begin{subfigure}{290\unitlength}
			\begin{picture}(290,290)
				\put(0,0){\includegraphics[trim=2.9cm 1.5cm 5.1cm 1.5cm, clip=true,width=290\unitlength]{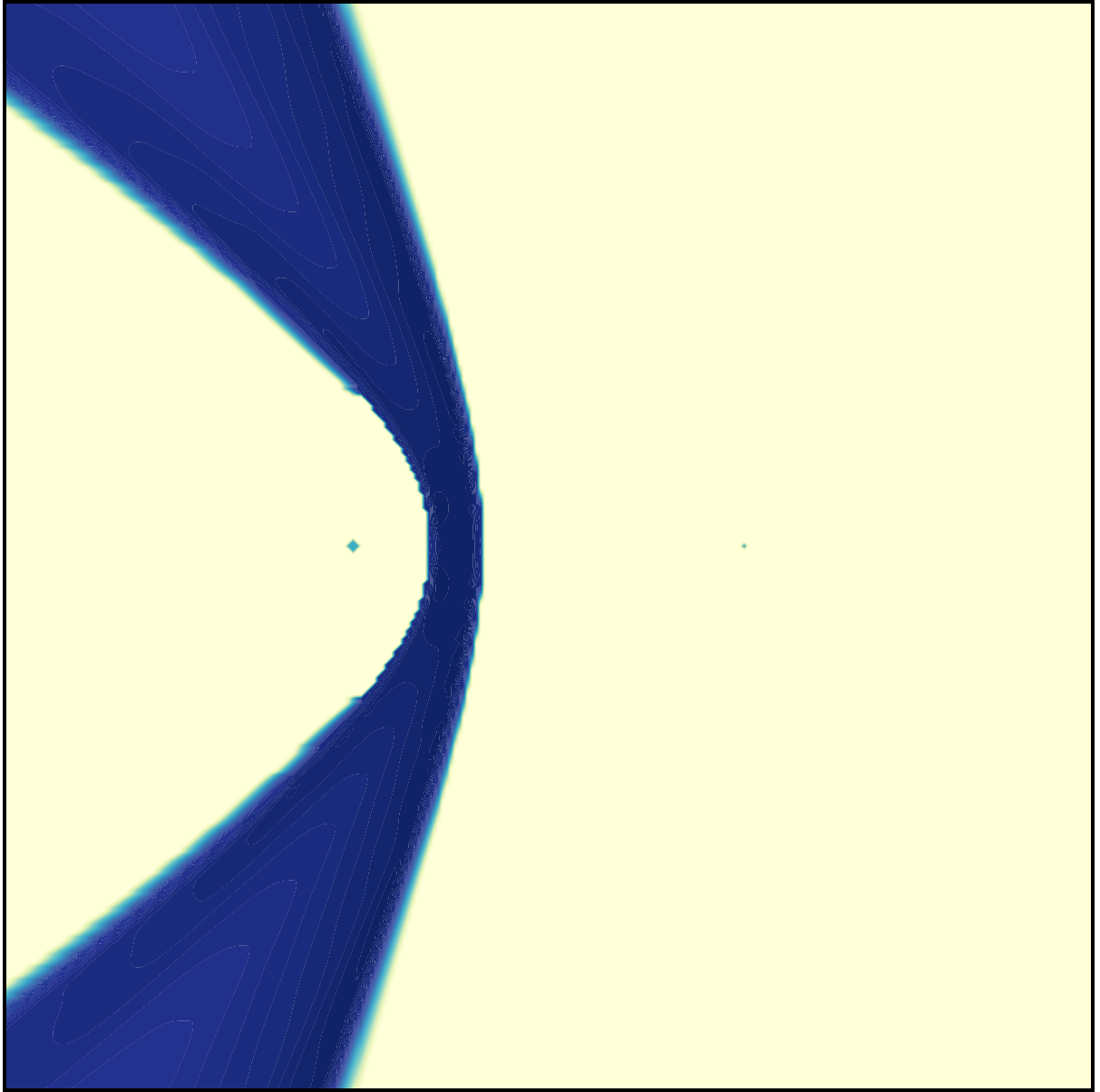}}
				\put(180,250){ 10 MeV}
				\put(126.7,20){\line(1,0){35.625}}
				\put(126.7,17){\line(0,1){6}}
				\put(162.325,17){\line(0,1){6}}
				\put(126.7,5){\scriptsize{500R$_\odot$}}
			\end{picture}
		\end{subfigure}
		\begin{subfigure}{290\unitlength}
			\begin{picture}(290,290)
				\put(0,0){\includegraphics[trim=2.9cm 1.5cm 5.1cm 1.5cm, clip=true,width=290\unitlength]{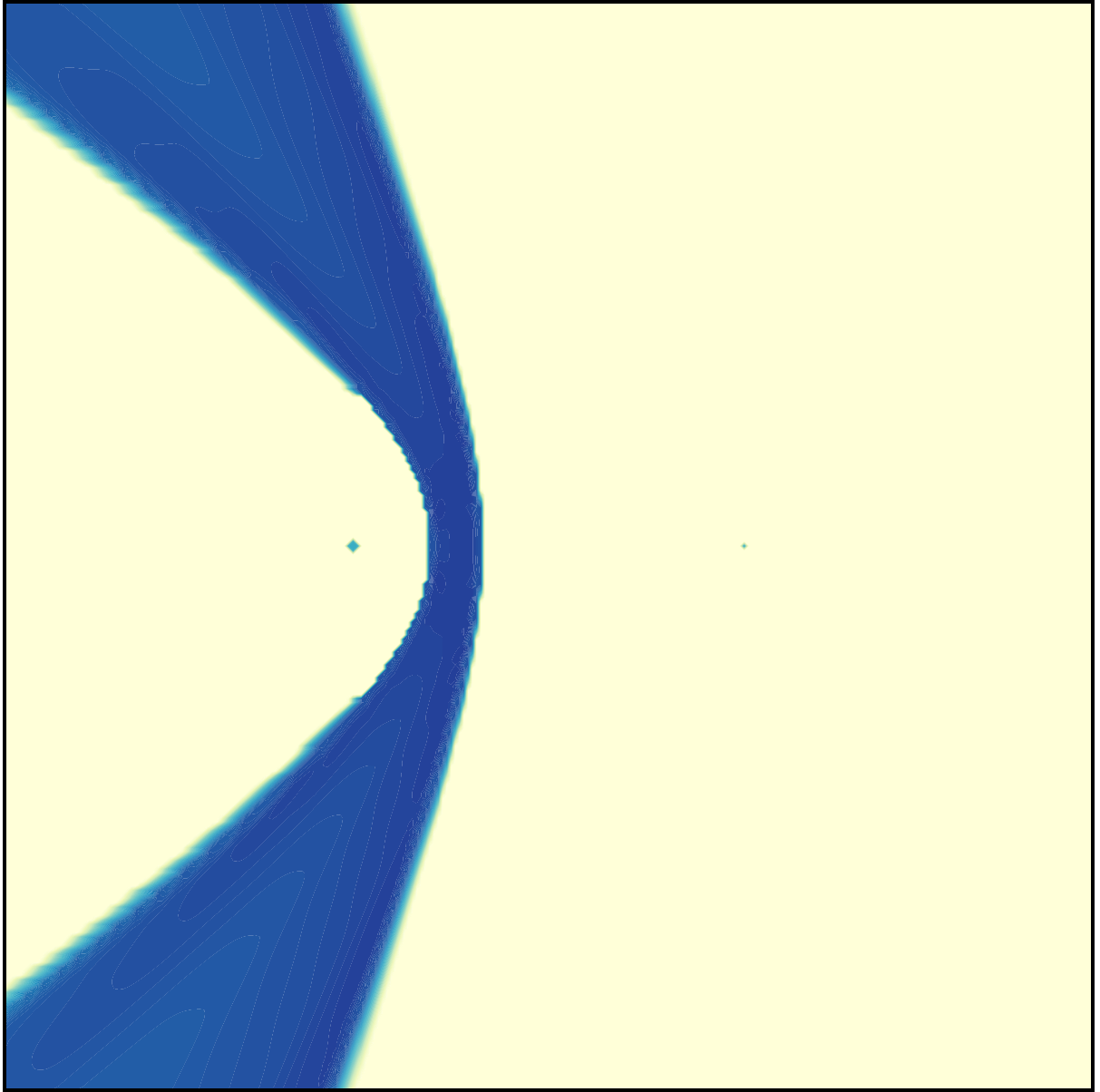}}
				\put(180,250){100 MeV}
				\put(126.7,20){\line(1,0){35.625}}
				\put(126.7,17){\line(0,1){6}}
				\put(162.325,17){\line(0,1){6}}
				\put(126.7,5){\scriptsize{500R$_\odot$}}
			\end{picture}
				\end{subfigure}
		\begin{subfigure}{290\unitlength}
			\begin{picture}(290,290)
				\put(0,0){\includegraphics[trim=2.9cm 1.5cm 5.1cm 1.5cm, clip=true,width=290\unitlength]{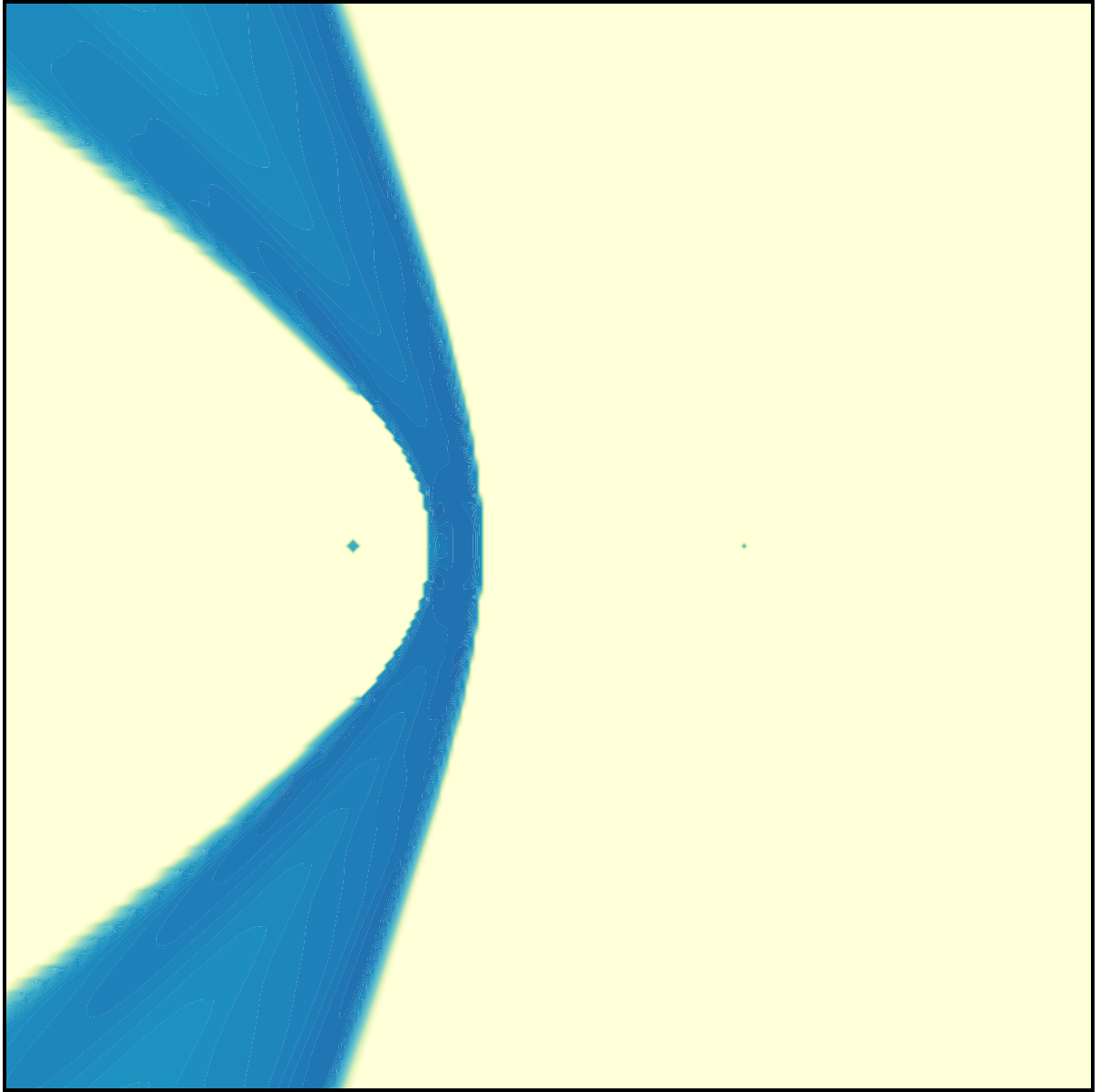}}
				\put(180,250){  1 GeV}
				\put(126.7,20){\line(1,0){35.625}}
				\put(126.7,17){\line(0,1){6}}
				\put(162.325,17){\line(0,1){6}}
				\put(126.7,5){\scriptsize{500R$_\odot$}}
			\end{picture}
		\end{subfigure}\\
		\begin{subfigure}{290\unitlength}
			\begin{picture}(290,290)
				\put(0,0){\includegraphics[trim=2.9cm 1.5cm 5.1cm 1.5cm, clip=true,width=290\unitlength]{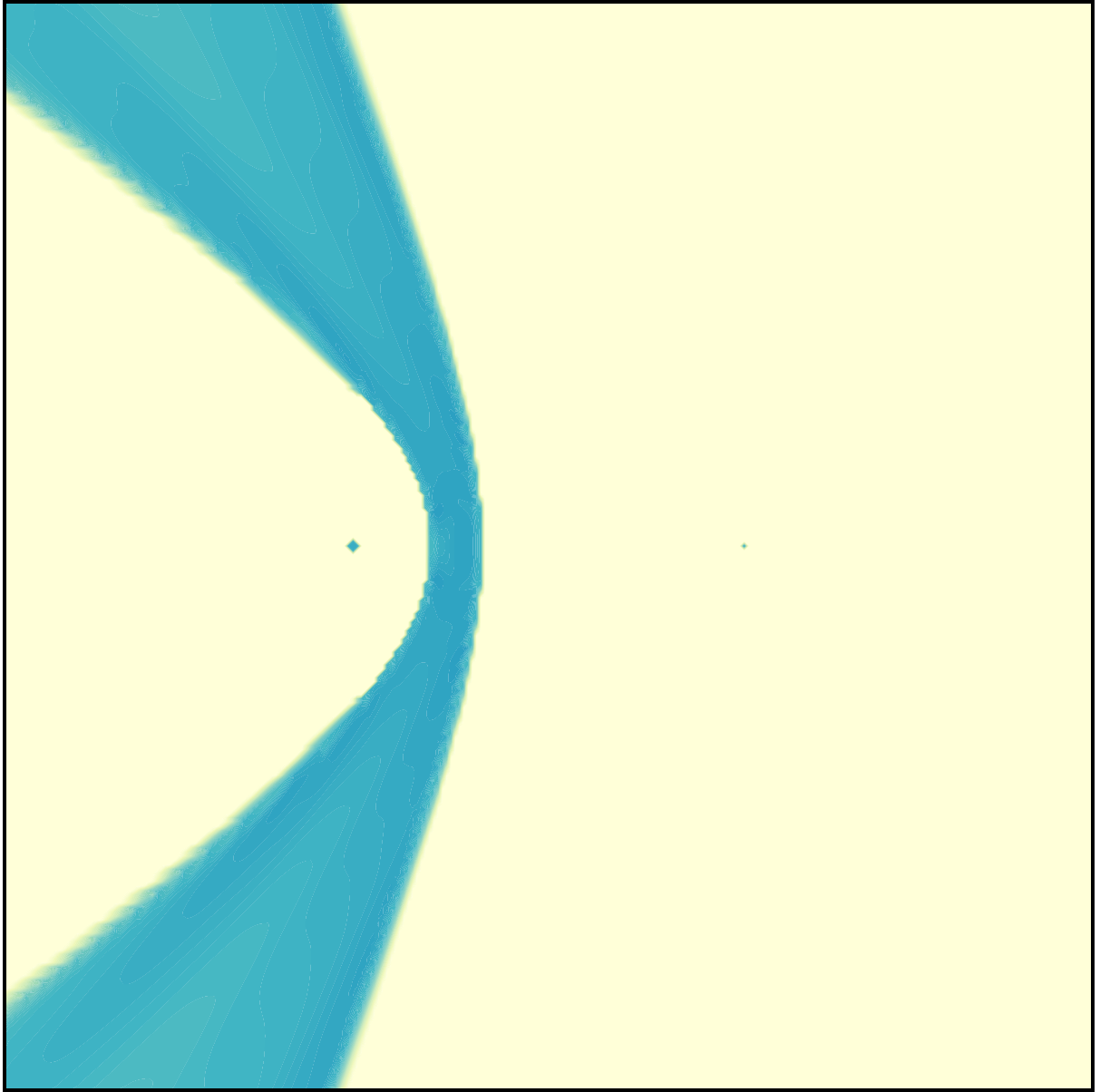}}
				\put(180,250){ 10 GeV}
				\put(126.7,20){\line(1,0){35.625}}
				\put(126.7,17){\line(0,1){6}}
				\put(162.325,17){\line(0,1){6}}
				\put(126.7,5){\scriptsize{500R$_\odot$}}
			\end{picture}
		\end{subfigure}
		\begin{subfigure}{290\unitlength}
			\begin{picture}(290,290)
				\put(0,0){\includegraphics[trim=2.9cm 1.5cm 5.1cm 1.5cm, clip=true,width=290\unitlength]{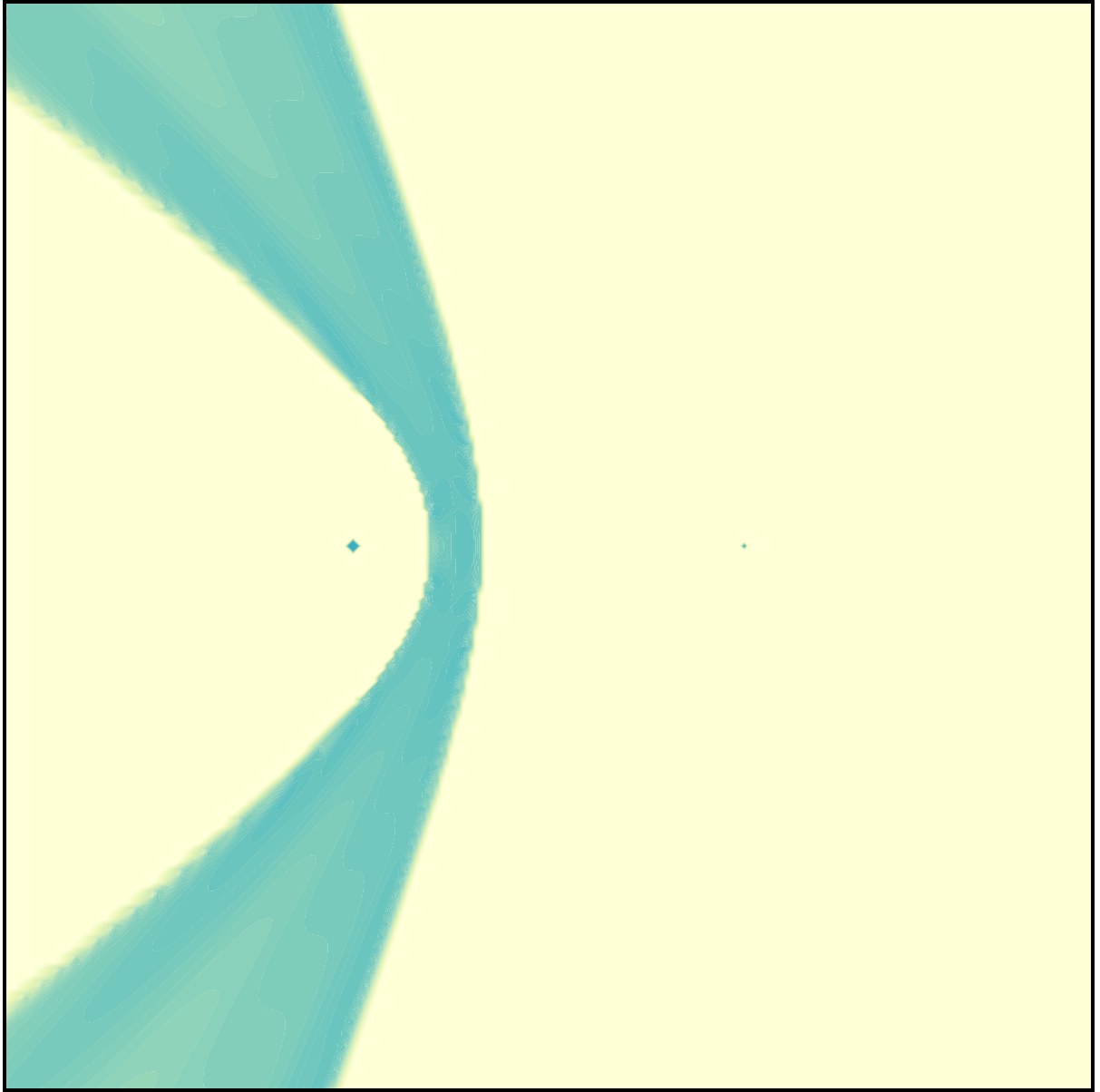}}
				\put(180,250){100 GeV}
				\put(126.7,20){\line(1,0){35.625}}
				\put(126.7,17){\line(0,1){6}}
				\put(162.325,17){\line(0,1){6}}
				\put(126.7,5){\scriptsize{500R$_\odot$}}
			\end{picture}
		\end{subfigure}
		\begin{subfigure}{290\unitlength}	
			\begin{picture}(290,290)
				\put(0,0){\includegraphics[trim=2.9cm 1.5cm 5.1cm 1.5cm, clip=true,width=290\unitlength]{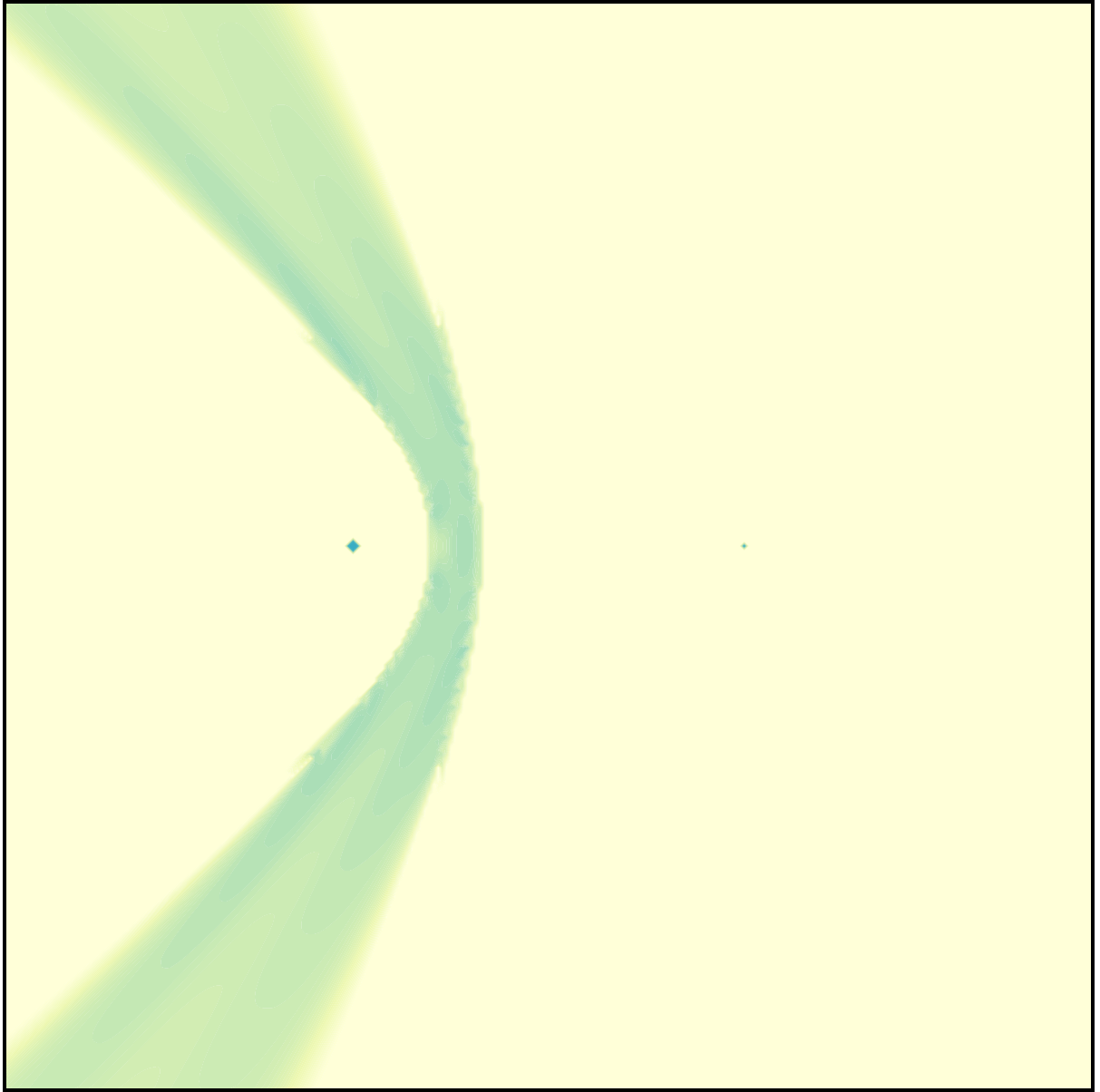}}
				\put(180,250){  1 TeV}
				\put(126.7,20){\line(1,0){35.625}}
				\put(126.7,17){\line(0,1){6}}
				\put(162.325,17){\line(0,1){6}}
				\put(126.7,5){\scriptsize{500R$_\odot$}}
			\end{picture}
		\end{subfigure}
	\end{subfigure}
	\begin{subfigure}{85\unitlength}
		\begin{picture}(85,580)
			\put(0,0){\includegraphics[trim=15.9cm 1.4cm 2.6cm 1.5cm, clip=true,height=580\unitlength]{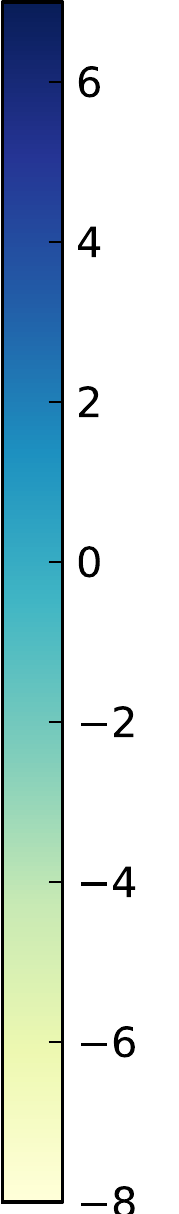}}
			\put(-24,200){\rotatebox{90}{log(MeV$^{-1}$m$^{-3}$)}}
		\end{picture}
	\end{subfigure}
\caption{Same as Fig. 4 for the case of protons.\label{prs}}
\end{figure}

\begin{figure}
	\setlength{\unitlength}{0.001\textwidth}
	\begin{subfigure}[c]{500\unitlength}
		\begin{picture}(500,370)
			\put(15,15){
				\includegraphics[trim=0.25cm 0.35cm 0.2cm 0cm,
				 clip=true,width=\textwidth]{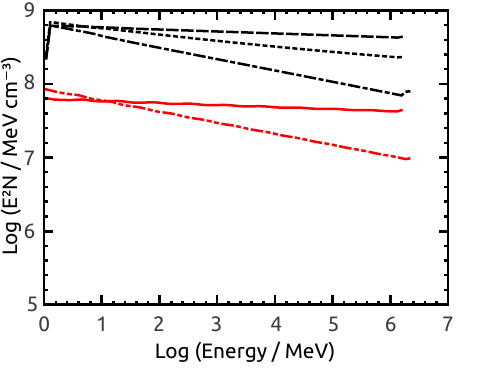}
			}
			\put(200,0){\footnotesize{log( E in MeV)}}
			\put(0,120){\rotatebox{90}{\footnotesize{log( $E^2N$ in MeV cm$^{-3}$ )}}}
			\put(440,80){a)}
		\end{picture}
	\end{subfigure}
	\begin{subfigure}[c]{500\unitlength}
		\begin{picture}(500,370)
			\put(15,15){
				\includegraphics[trim=0.25cm 0.35cm 0.2cm 0cm,
				 clip=true,width=\textwidth]{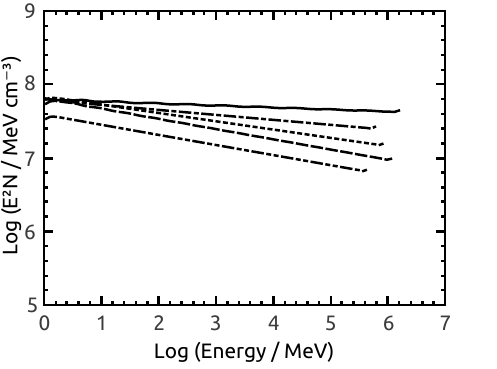}
			}
			\put(200,0){\footnotesize{log( E in MeV)}}
			\put(0,120){\rotatebox{90}{\footnotesize{log( $E^2N$ in MeV cm$^{-3}$ )}}}
			\put(440,80){b)}
		\end{picture}
	\end{subfigure}
	\begin{subfigure}[c]{500\unitlength}
		\begin{picture}(500,400)
			\put(15,15){
			\includegraphics[trim=0.25cm 0.35cm 0.2cm 0cm,
				 clip=true,width=\textwidth]{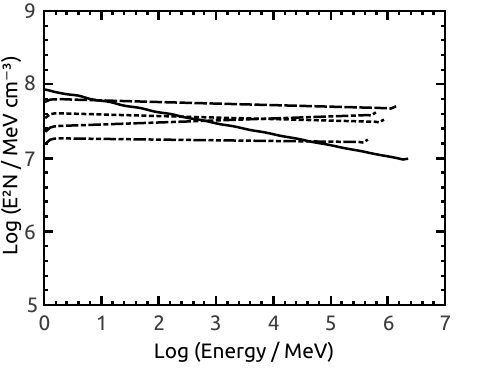}
			}
			\put(200,0){\footnotesize{log( E in MeV )}}
			\put(0,120){\rotatebox{90}{\footnotesize{log( $E^2N$ in MeV cm$^{-3}$ )}}}
			\put(440,80){c)}
		\end{picture}
	\end{subfigure}
	\begin{subfigure}[c]{500\unitlength}
		\begin{picture}(500,400)
			\put(15,15){
			\includegraphics[trim=0.25cm 0.35cm 0.2cm 0cm,
				 clip=true,width=\textwidth]{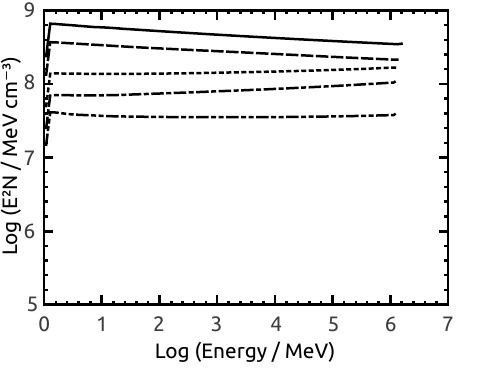}
				}
			\put(200,0){\footnotesize{log( E in MeV )}}
			\put(0,120){\rotatebox{90}{\footnotesize{log( $E^2N$ in MeV cm$^{-3}$ )}}}
			\put(440,80){d)}
		\end{picture}	
			\end{subfigure}
			\caption{Proton spectra for various positions within the WCR. Fig. (a) shows spectra along the connecting line of the stars, where distance from the shock front facing the WR wind (solid) is $\sim$60 (dashed), $\sim$110 (dotted), $\sim$160 (dash-dotted) and $\sim$190 R$_\odot$ (double-dot--dashed). Spectra within the acceleration region are coloured in red. Fig. (b) to (d) show regions along the WR shock (b), along the B shock (c) and along the centre of the WCR (d). Here, the distance to the corresponding region at the apex (solid) is  $\sim$500 (dashed), $\sim$1000 (dotted), $\sim$1500 (dash-dotted) and $\sim$2000 R$_\odot$ (double-dot--dashed).  \label{prspec} }
\end{figure}

\begin{figure}
	\setlength{\unitlength}{0.001\textwidth}
	\begin{subfigure}[l]{900\unitlength}
		\begin{subfigure}{290\unitlength}
			\begin{picture}(290,290)
				\put(0,0){\includegraphics[trim=2.9cm 1.5cm 5.1cm 1.5cm, clip=true,width=290\unitlength]{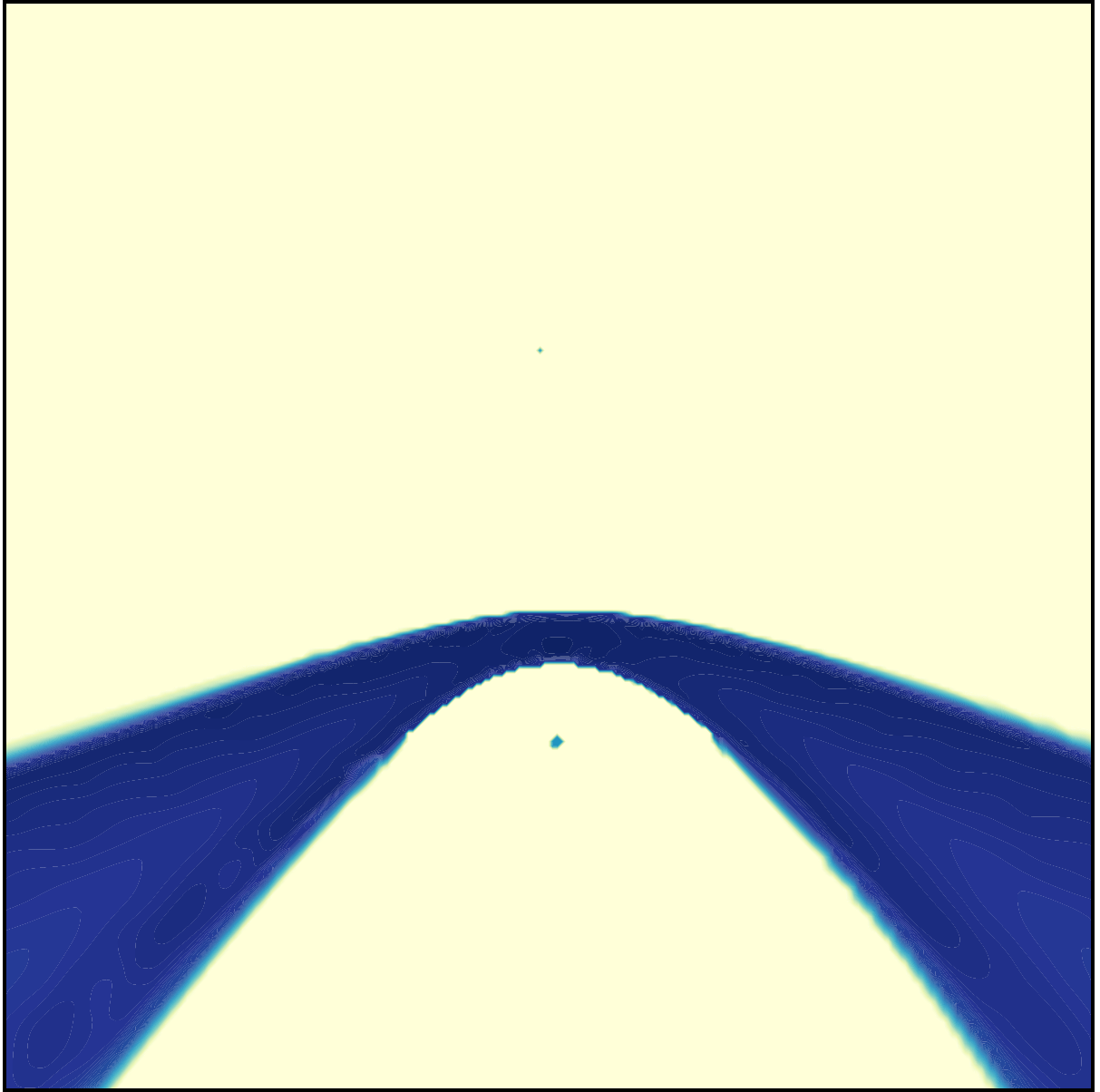}}
				\put(180,250){ 10 MeV}
				\put(126.7,20){\line(1,0){35.625}}
				\put(126.7,17){\line(0,1){6}}
				\put(162.325,17){\line(0,1){6}}
				\put(126.7,5){\scriptsize{500R$_\odot$}}
			\end{picture}
		\end{subfigure}
		\begin{subfigure}{290\unitlength}
			\begin{picture}(290,290)
				\put(0,0){\includegraphics[trim=2.9cm 1.5cm 5.1cm 1.5cm, clip=true,width=290\unitlength]{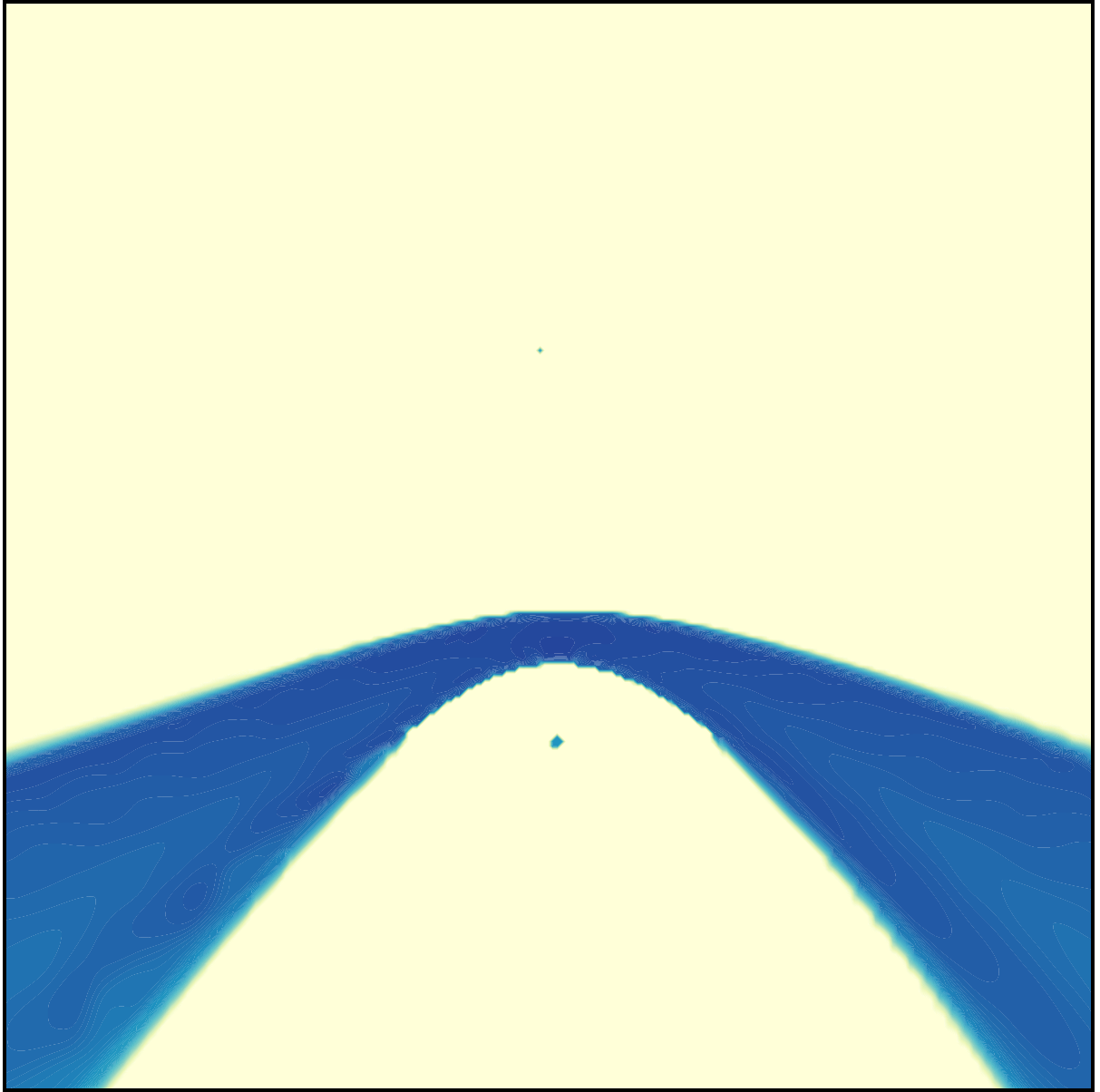}}
				\put(180,250){100 MeV}
				\put(126.7,20){\line(1,0){35.625}}
				\put(126.7,17){\line(0,1){6}}
				\put(162.325,17){\line(0,1){6}}
				\put(126.7,5){\scriptsize{500R$_\odot$}}
			\end{picture}
				\end{subfigure}
		\begin{subfigure}{290\unitlength}
			\begin{picture}(290,290)
				\put(0,0){\includegraphics[trim=2.9cm 1.5cm 5.1cm 1.5cm, clip=true,width=290\unitlength]{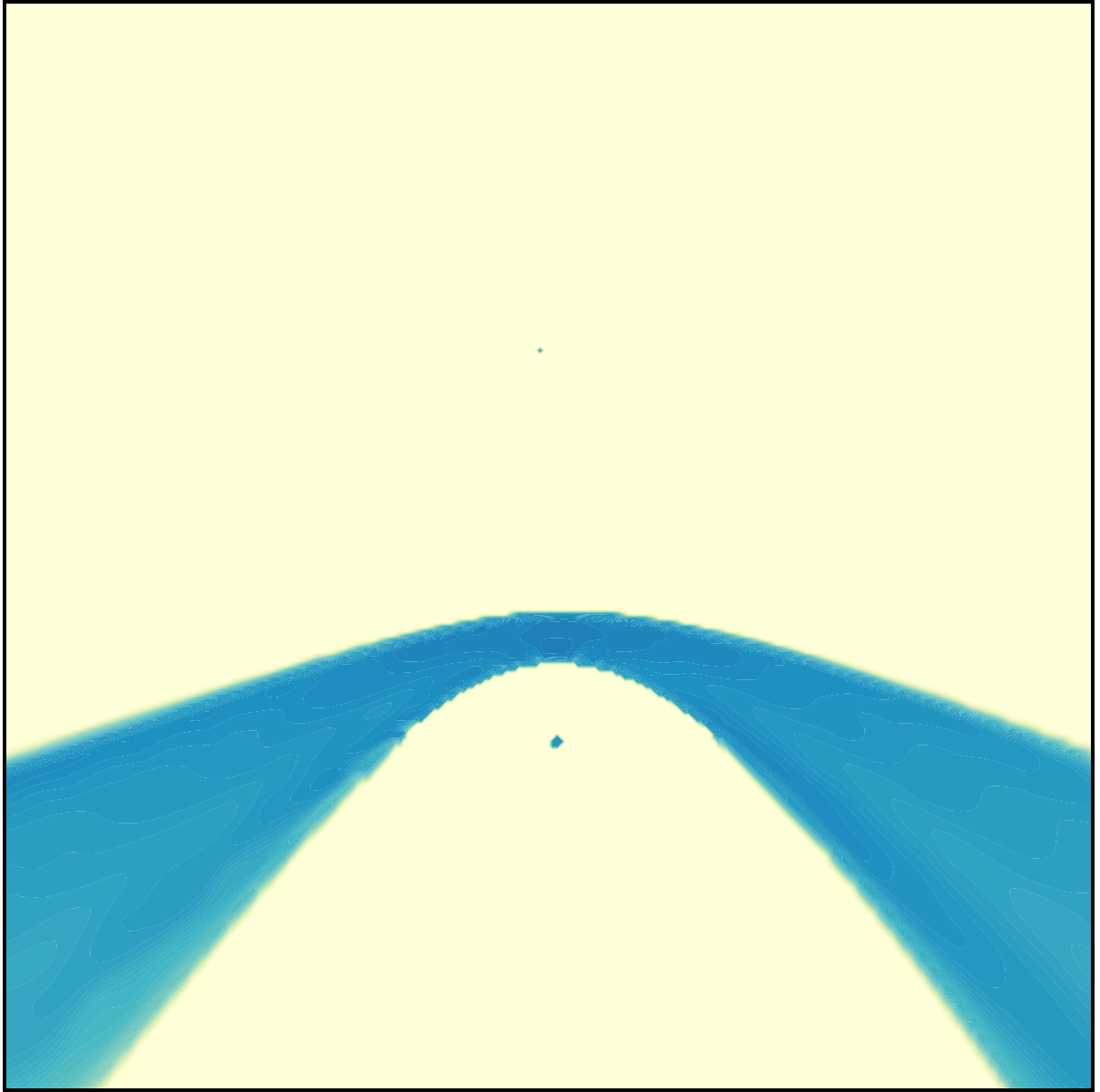}}
				\put(180,250){  1 GeV}
				\put(126.7,20){\line(1,0){35.625}}
				\put(126.7,17){\line(0,1){6}}
				\put(162.325,17){\line(0,1){6}}
				\put(126.7,5){\scriptsize{500R$_\odot$}}
			\end{picture}
		\end{subfigure}\\
		\begin{subfigure}{290\unitlength}
			\begin{picture}(290,290)
				\put(0,0){\includegraphics[trim=2.9cm 1.5cm 5.1cm 1.5cm, clip=true,width=290\unitlength]{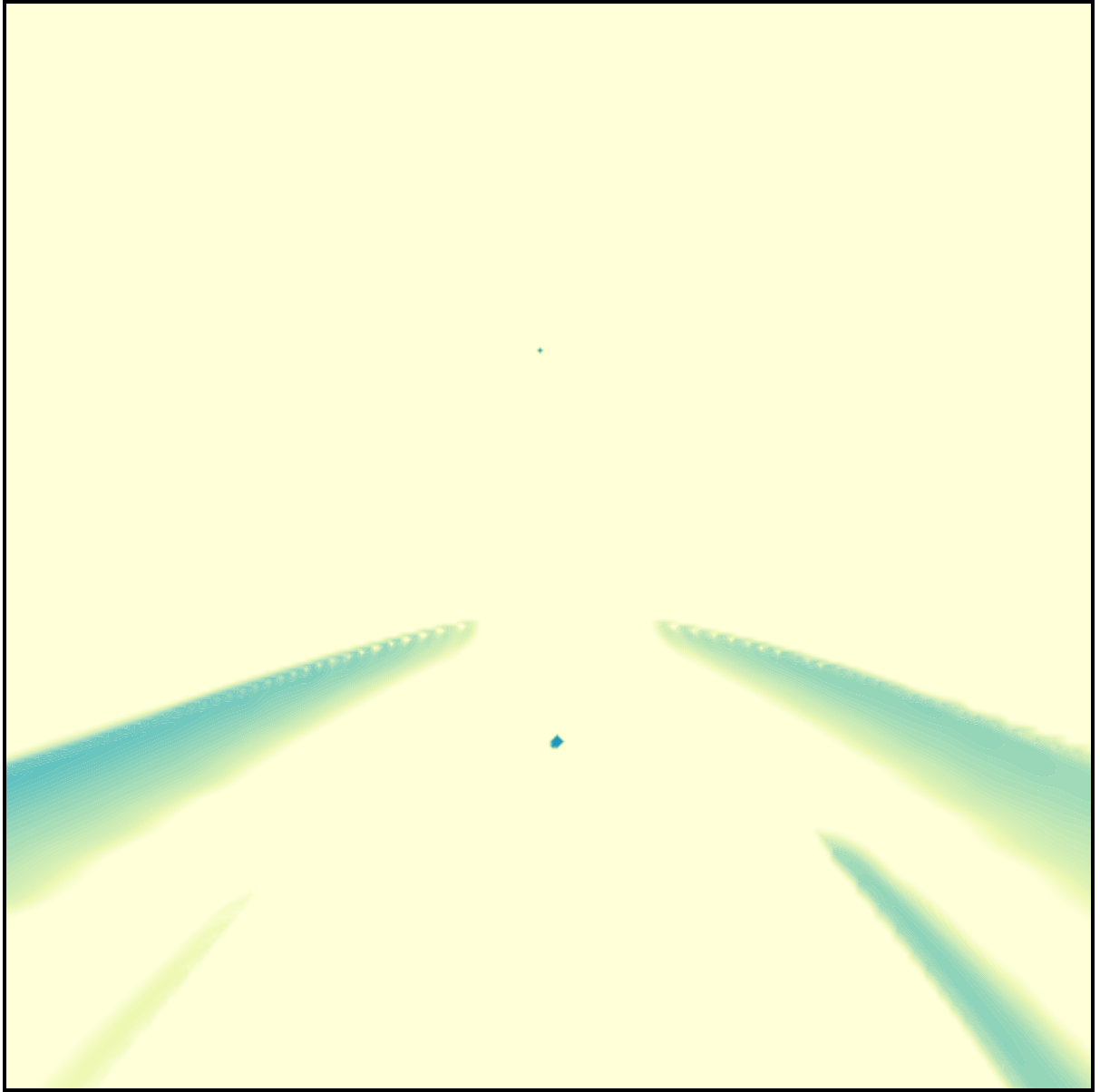}}
				\put(180,250){ 10 GeV}
				\put(126.7,20){\line(1,0){35.625}}
				\put(126.7,17){\line(0,1){6}}
				\put(162.325,17){\line(0,1){6}}
				\put(126.7,5){\scriptsize{500R$_\odot$}}
			\end{picture}
		\end{subfigure}
		\begin{subfigure}{290\unitlength}
			\begin{picture}(290,290)
				\put(0,0){\includegraphics[trim=2.9cm 1.5cm 5.1cm 1.5cm, clip=true,width=290\unitlength]{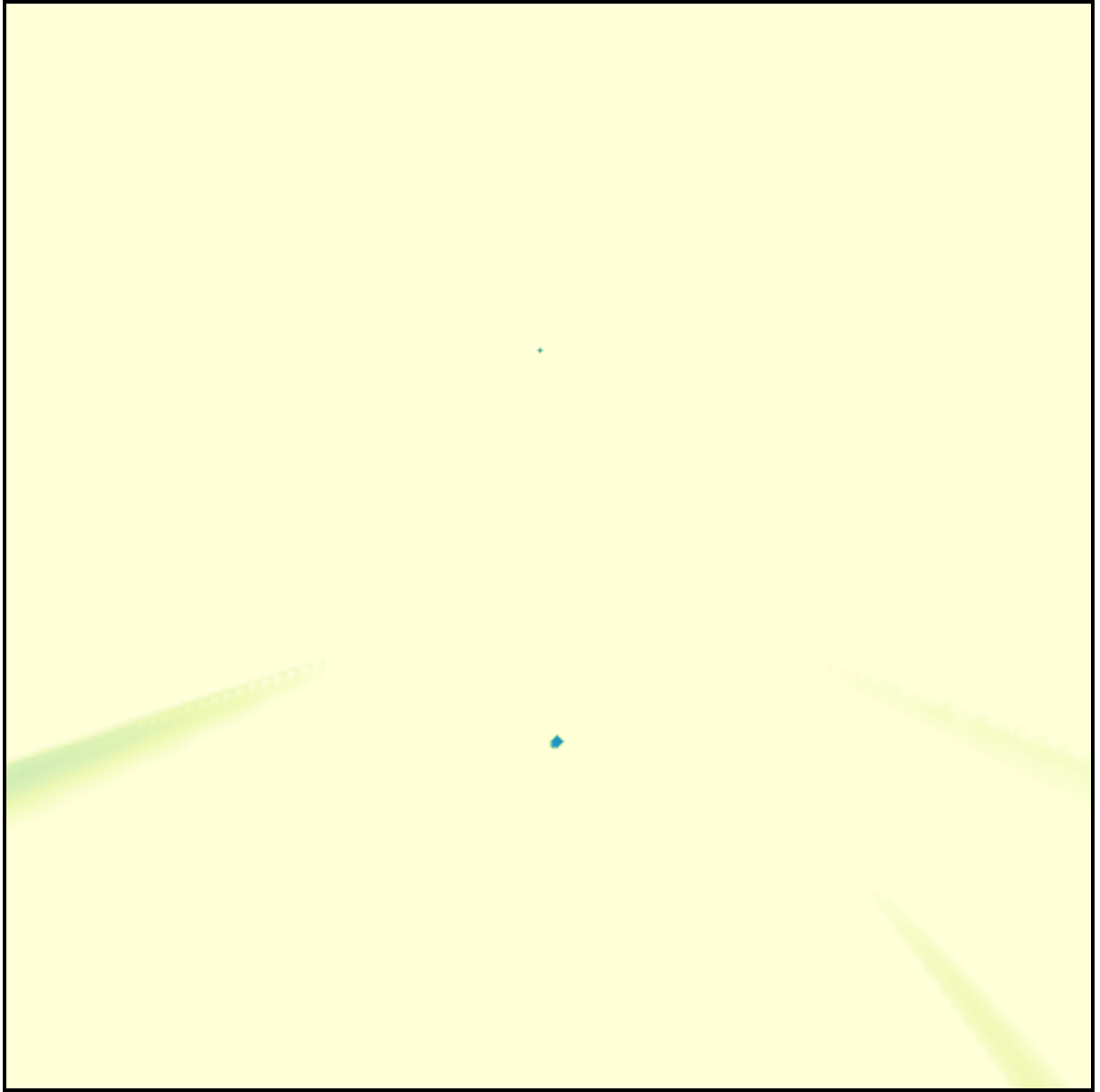}}
				\put(180,250){100 GeV}
				\put(126.7,20){\line(1,0){35.625}}
				\put(126.7,17){\line(0,1){6}}
				\put(162.325,17){\line(0,1){6}}
				\put(126.7,5){\scriptsize{500R$_\odot$}}
			\end{picture}
		\end{subfigure}
		\begin{subfigure}{290\unitlength}	
			\begin{picture}(290,290)
				\put(0,0){\includegraphics[trim=2.9cm 1.5cm 5.1cm 1.5cm, clip=true,width=290\unitlength]{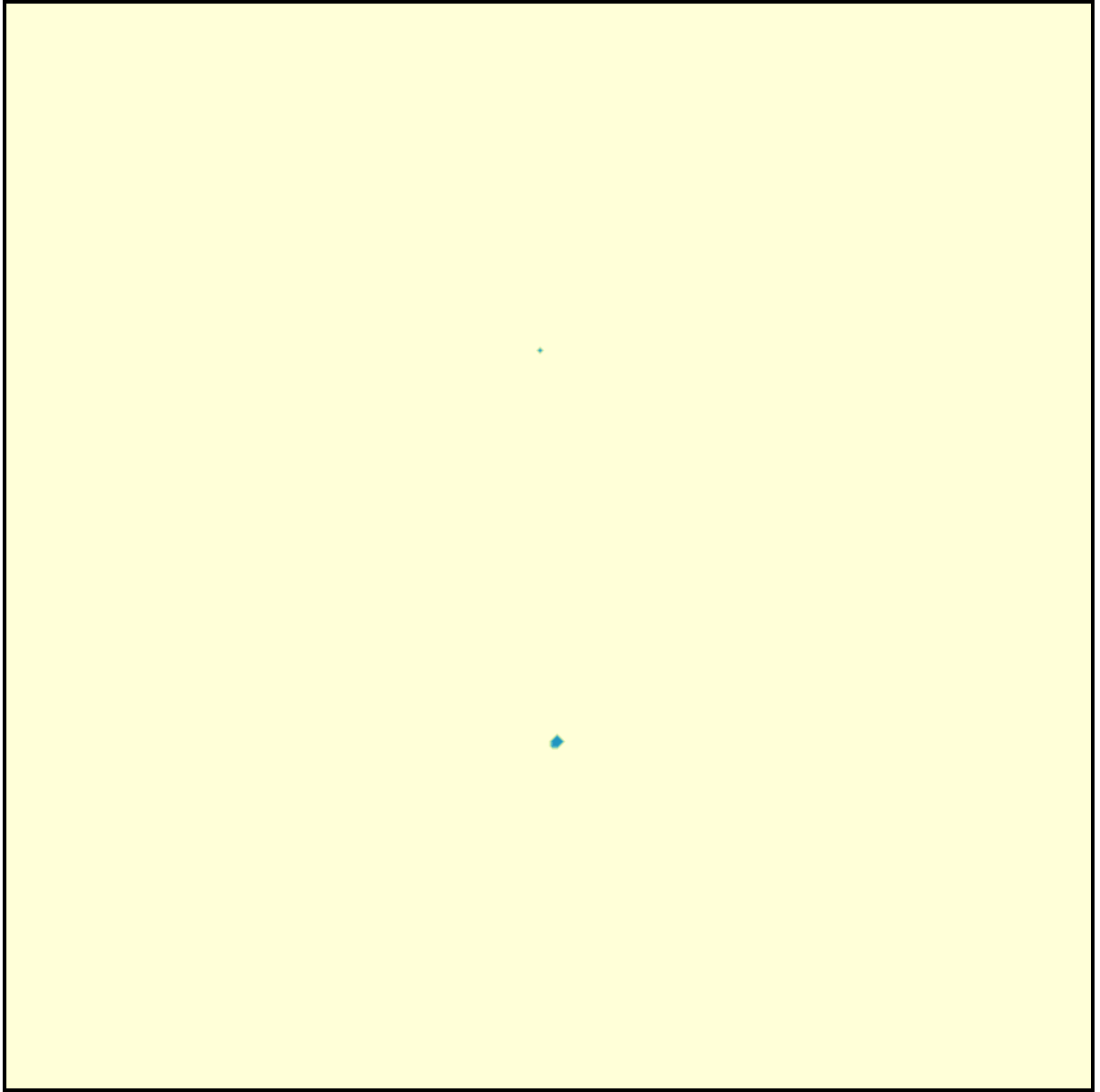}}
				\put(180,250){  1 TeV}
				\put(126.7,20){\line(1,0){35.625}}
				\put(126.7,17){\line(0,1){6}}
				\put(162.325,17){\line(0,1){6}}
				\put(126.7,5){\scriptsize{500R$_\odot$}}
			\end{picture}
		\end{subfigure}
	\end{subfigure}
	\begin{subfigure}{85\unitlength}
		\begin{picture}(85,580)
			\put(0,0){\includegraphics[trim=15.9cm 1.5cm 2.6cm 1.5cm, clip=true,height=580\unitlength]{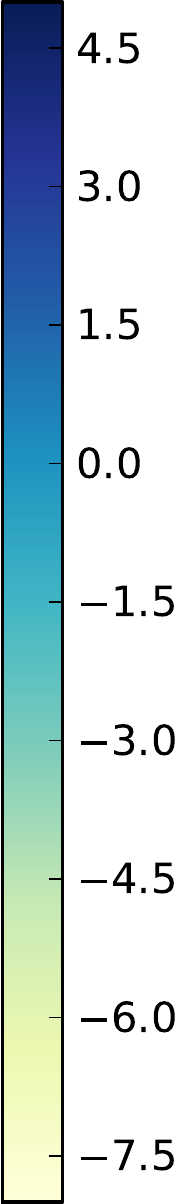}}
			\put(-24,200){\rotatebox{90}{log(MeV$^{-1}$m$^{-3}$)}}
		\end{picture}
	\end{subfigure}
	\caption{Same as Fig. 4, but with orbital motion. The stars move counter-clockwise on a circular orbit. \label{elsrot}}
\end{figure}
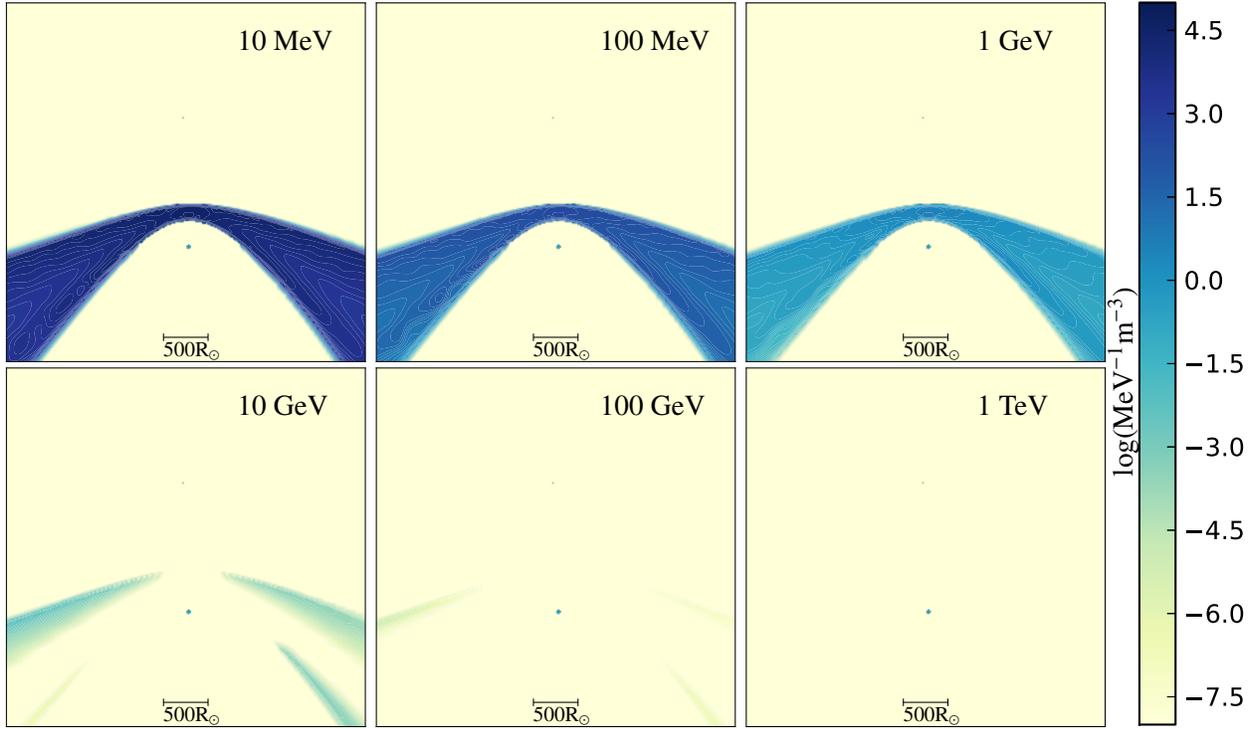

\begin{figure}
	\setlength{\unitlength}{0.001\textwidth}
	\begin{subfigure}[c]{500\unitlength}
		\begin{picture}(500,400)
			\put(15,15){
				\includegraphics[trim=0.25cm 0.35cm 0.2cm 0cm,
				 clip=true,width=\textwidth]{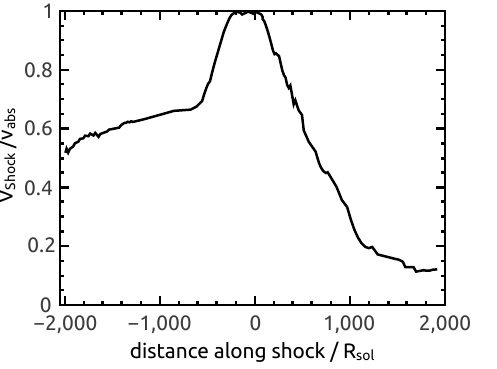}
			}
			\put(170,0){\footnotesize{distance along shock in MeV}}
			\put(0,160){\rotatebox{90}{\footnotesize{$V_\mathrm{Shock}/v_\mathrm{abs}$}}}
			\put(440,60){a)}			
		\end{picture}
	\end{subfigure}
	\begin{subfigure}[c]{500\unitlength}
		\begin{picture}(500,400)
			\put(15,15){
				\includegraphics[trim=0.32cm 0.35cm 0.14cm 0cm,
				 clip=true,width=\textwidth]{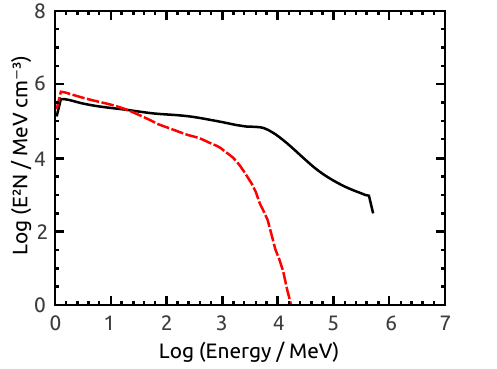}
			}
			\put(170,0){\footnotesize{Log( E in MeV )}}
			\put(0,120){\rotatebox{90}{\footnotesize{log( $E^2N$ in MeV cm$^{-3}$ )}}}
			\put(440,60){b)}
		\end{picture}
	\end{subfigure}
			\caption{a) Ratio of the shock normal velocity to overall wind velocity as a function of distance from the apex along the B side of the WCR shock front. Negative distance traverses the shock along the forward arm, positive distance traverses the shock along the trailing arm of the WCR.\\
b) Electron spectra for two characteristic positions in the forward arm (black solid) and in the trailing arm (red dashed) of the WCR close to the shock towards the B star. \label{orbitspec} }

\end{figure}
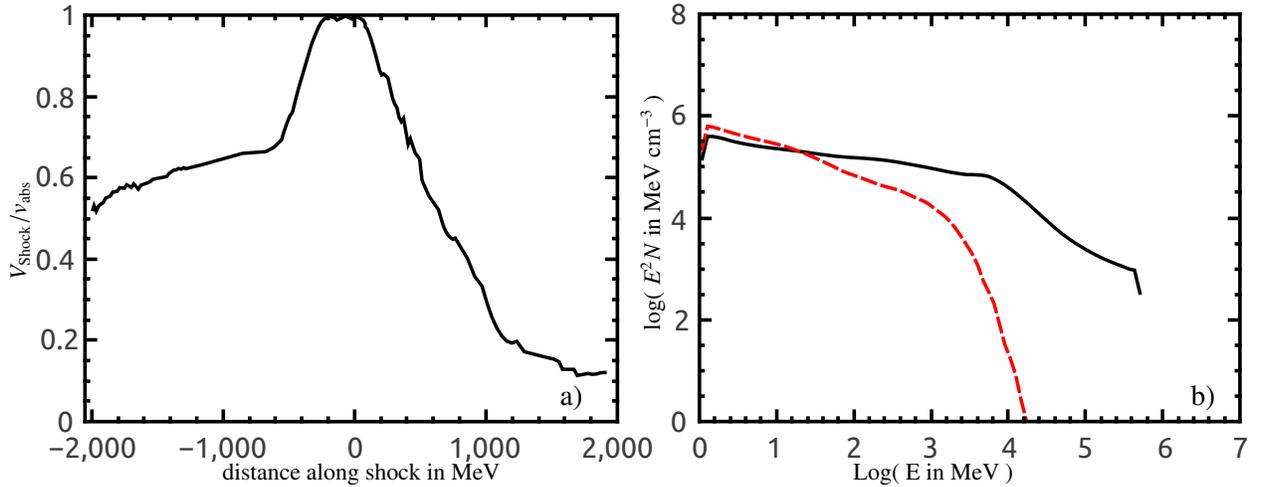

\end{document}